\documentclass[fleqn,10pt]{wlscirep}
\usepackage[utf8]{inputenc}
\usepackage[T1]{fontenc}
\usepackage{bm}
\usepackage{graphicx}
\usepackage{xcolor}
\usepackage{lineno}
\usepackage{amssymb}

\newcommand{\red}[1]{\textcolor{black}{#1}}

\def\K212{K$_2$Co(SeO$_3$)$_2$}
\DeclareMathAlphabet\mathbfcal{OMS}{cmsy}{b}{n}

\title{Phase Diagram and Spectroscopic Signatures of a
Supersolid in the Quantum Ising Magnet \K212}

\author[1,11,*]{Tong Chen}
\author[1,11]{Alireza Ghasemi}
\author[1]{Junyi Zhang}
\author[1]{Liyu Shi}
\author[1]{Zhenisbek Tagay}
\author[1]{Youzhe Chen}
\author[2]{Lei Chen}
\author[3]{Eun-Sang Choi}
\author[4,10]{Marcelo Jaime}
\author[4]{Minseong Lee}
\author[5]{Yiqing Hao}
\author[5]{Huibo Cao}
\author[5]{Barry Winn}
\author[5]{Andrey A. Podlesnyak}
\author[5]{Daniel M. Pajerowski}
\author[6,7,*]{Ruidan Zhong}
\author[7]{Xianghan Xu}
\author[1]{N. P. Armitage}
\author[7]{Robert Cava}
\author[1,8,9,*]{Collin Broholm}

\affil[1]{Institute for Quantum Matter and Department of Physics and Astronomy, Johns Hopkins University, Baltimore, Maryland, USA.}
\affil[2]{Department of Physics and Astronomy, Rice University, Houston, Texas, USA.}
\affil[3]{National High Magnetic Field Laboratory and Department of Physics, Florida State University, Tallahassee, USA}
\affil[4]{National High Magnetic Field Laboratory, Los Alamos National Laboratory, Los Alamos, New Mexico, USA.}
\affil[5]{Neutron Scattering Division, Oak Ridge National Laboratory, Oak Ridge, Tennessee, USA.}
\affil[6]{Tsung-Dao Lee Institute and School of Physics and Astronomy, Shanghai Jiao Tong University, Shanghai, China.}
\affil[7]{Department  of  Chemistry,  Princeton  University,  Princeton,  New Jersey, USA.}
\affil[8]{NIST Center for Neutron Research, Gaithersburg, Maryland, USA.}
\affil[9]{Department of Materials Science and Engineering, Johns Hopkins University, Baltimore, Maryland, USA.}
\affil[10]{New address: Physikalisch-Technische Bundesanstalt, Braunschweig, Germany}
\affil[11]{These authors contributed equally: Tong Chen, Alireza Ghasemi.}

\affil[*]{e-mail: tchen115@jhu.edu, rzhong@sjtu.edu.cn, broholm@jhu.edu}

\begin{abstract}
A supersolid is a quantum-entangled state of matter exhibiting the dual characteristics of superfluidity and solidity. Theory predicts that hard-core bosons with repulsive interactions on a triangular lattice can form supersolid phases at half filling and near complete filling. Leveraging an exact mapping between bosons and spin-$\frac{1}{2}$ degrees of freedom, we investigate these phases in the spin-$\frac{1}{2}$ triangular-lattice antiferromagnet \K212 with exchange constants $J_z = 2.96(2)$~meV and $J_{\perp} = 0.21(3)$~meV. At zero field, neutron diffraction reveals the gradual development for $T<15$~K of quasi-two-dimensional $\sqrt{3}\times\sqrt{3}$ magnetic order with $Z_3$ translational symmetry breaking (solidity) albeit with 44(5)\% reduced amplitude at $T=0.3$~K indicating strong quantum fluctuations. These are apparent in equidistant bands of continuum neutron scattering for $\hslash\omega_n\approx n\times J_z$, where $n=0,1,2,3$. The lowest energy ($n=0$) $\bf Q$-dependent continuum has a lower resonant edge and includes a quasi-elastic component at K $(\frac{1}{3}\frac{1}{3})$ consistent with broken $U(1)$ spin rotational symmetry (boson superfluidity). Competing instabilities are apparent in soft albeit finite-energy modes at M $(\frac{1}{2}0)$  and at $\frac{1}{2}$K $(\frac{1}{6}\frac{1}{6})$. For $\bf c$-axis-oriented magnetic fields $17~{\rm T} <\mu_0 H< 21~{\rm T}$ that almost saturate the magnetization, corresponding to nearly filling the lattice with bosons, we find a new phase consistent with a second supersolid. These phases are separated by a pronounced 1/3 magnetization plateau that supports coherent spin waves, from which we determine the spin Hamiltonian.
\end{abstract}
\begin{document}

\flushbottom
\maketitle

\thispagestyle{empty}

\section*{Introduction}

Frustrated magnets can host exotic states of matter as the macroscopic degeneracies resulting from competing interactions are lifted by quantum fluctuations. In three dimensions, for example, Ising spins with ferromagnetic nearest-neighbor interactions on the pyrochlore lattice of corner-sharing tetrahedra form a degenerate ``spin ice'' manifold with a residual Pauling entropy $S(T=0)\approx \frac{1}{2} \log(\frac{3}{2})R\approx0.203R$. \cite{ramirez1999zero, bramwell2001spin} Theory indicates that from this manifold, quantum fluctuations can generate a quantum spin ice phase with quasi-particles sourcing the fields of emergent electromagnetism. \cite{savary2012coulombic, lee2012generic, gingras2014quantum}

In two dimensions, Ising spins with antiferromagnetic interactions on a triangular lattice form a degenerate manifold with an entropy of $S(0)=0.323R$ determined by Wannier.\cite{wannier1950antiferromagnetism} Recent theoretical work indicates that adding quantum fluctuations to this manifold can produce a supersolid \cite{jiang2009supersolid,heidarian2010supersolidity,yamamoto2014quantum,sellmann2015phase} where superfluidity and solidity coexist.\cite{thouless1969flow, andreev1969quantum, chester1970speculations, leggett1970can} Originally studied in the context of solid $^4$He, \cite{kim2004probable, boninsegni2012colloquium} the concept of supersolidity has been extended to ultracold gases \cite{li2017stripe, leonard2017supersolid} and triangular lattices of hard-core bosons,\cite{melko2005supersolid, wessel2005supersolid, boninsegni2005supersolid, heidarian2005persistent, wang2009extended} which map to the spin-$\frac{1}{2}$ Ising model with $|\pm\frac{1}{2}\rangle$ spin states representing the presence and absence of a boson, respectively (see Methods).

The experimental realization of emergent quantum many-body phases like these in frustrated magnets must contend with subleading interactions\cite{den2000dipolar} and chemical disorder,\cite{syzranov2022eminuscent} which inevitably rival thermal \cite{villain1980order, henley1989ordering} and quantum fluctuations in lifting degeneracies. Suitable model systems are further constrained by the availability of spectroscopic tools with sufficient resolution and sensitivity \cite{stone2014comparison} to characterize the emergent phases.\cite{chamorro2020chemistry} To date, candidate materials for quantum spin ice are based on tri-valent magnetic rare-earth elements\cite{gardner2010magnetic} where dominant interactions are on the order of 1 K, leading to emergent energy scales in the mK range. This approaches the limits of instrumental resolution and precludes detailed experimental exploration of the emergent properties, though evidence for a $\pi$-flux quantum spin ice phase in $\rm Ce_2Zr_2O_7$ is mounting. \cite{gao2019experimental,gaudet2019quantum}

In contrast, hexagonal di-valent transition metal oxides can form quasi-two-dimensional magnets with an order of magnitude stronger interactions. Just as emergent properties of one-dimensional quantum magnets -- including the gapless spinon continuum of the spin-$\frac{1}{2}$ chain \cite{tennant1993unbound, lake2005quantum, mourigal2013fractional} and the Haldane valence bond solid  \cite{ma1992dominance, zaliznyak2001continuum} -- were characterized in transition-metal-based magnets using inelastic neutron scattering, hexagonal oxides with well-separated layers of transition metal ions that carry a spin-orbital magnetic moment, \cite{liu2023non} provide a promising platform to explore the effects of quantum fluctuations on Wannier's manifold and to search for the predicted supersolids.


Recent work on ${\rm Na_2BaCo(PO_4)_2}$ has drawn attention to this area. However, while beneficial for cryogenic applications, \cite{xiang2024giant} the weak superexchange interactions mediated by the tetrahedral polyanion $\rm PO_4$ \cite{popescu2025zeeman} preclude detailed neutron spectroscopy. In addition, the interlayer interactions are strong enough to induce conventional magnetic order before the emergent phase is fully developed.\cite{sheng2025continuum} Indeed, it now appears that the continuum scattering in ${\rm Na_2BaCo(PO_4)_2}$ is actually heterogeneous spin wave scattering resulting from the incommensurate inter-plane order.\cite{1pvl-kzjm} The situation is more favorable in \K212 (KCSO), where the triangular lattice of spin-$\frac{1}{2}$ Co$^{2+}$ is built around the $\rm SeO_3$ polyanion that mediates an order of magnitude stronger and more anisotropic interactions within the triangular lattice planes. \cite{zhong2020frustrated} The absence of a sharp thermal anomaly in the zero-field specific heat capacity $C_{\rm p}(T)$ down to the mK regime indicates a highly two-dimensional spin system. Here, we explore the emergent properties of KCSO through heat capacity, magnetization, and neutron scattering experiments. We show that it is accurately described by the following spin Hamiltonian
\begin{equation}
{\cal H} = \sum_{\langle i,j\rangle} [J_z  S_i^z S_j^z + J_{\perp} (S_i^x S_j^x + S_i^y S_j^y)] - g_z \mu_{\rm B} \mu_0 H_z \sum_i S_i^z.
\label{Eq: Hxxz}
\end{equation}
The first sum is over nearest neighbors on a triangular lattice, $J_z = 2.96(2)$~meV and $J_{\perp} = 0.21(3)$~meV, $S_k^\alpha$ is a spin-$\frac{1}{2}$ operator, and the second sum is the Zeeman term with $g_z=7.8$. 
In zero field we observe quasi-2D $\sqrt{3}\times\sqrt{3}$ antiferromagnetic correlations below $15$~K. For $T=0.1$~K, the 44(5)\% reduced root mean squared (RMS) ordered moment and the lack of ferromagnetic correlations over a similar length scale indicate quantum fluctuations. These take the form of four separate, bounded continua with roton-like minima in the lowest energy band signaling competing instabilities. At $T=0.29$~K, the coexistence of a Goldstone mode and a pseudo-Goldstone mode at 0.060(4) meV signals the breaking of $U(1)$ rotational symmetry and $Z_3$ translational symmetry -- the defining characteristics of a supersolid. By mapping the field-temperature phase diagram, we identify a phase transition into a 1/3 magnetization plateau belonging to the two-dimensional (2D) 3-state Potts universality class. We also present experimental evidence for a distinct high-field phase immediately below full saturation, consistent with a second theoretically predicted supersolid phase.

\section*{Magnetic Order}

\begin{figure}[ht]
\centering
\includegraphics[width=1\textwidth]{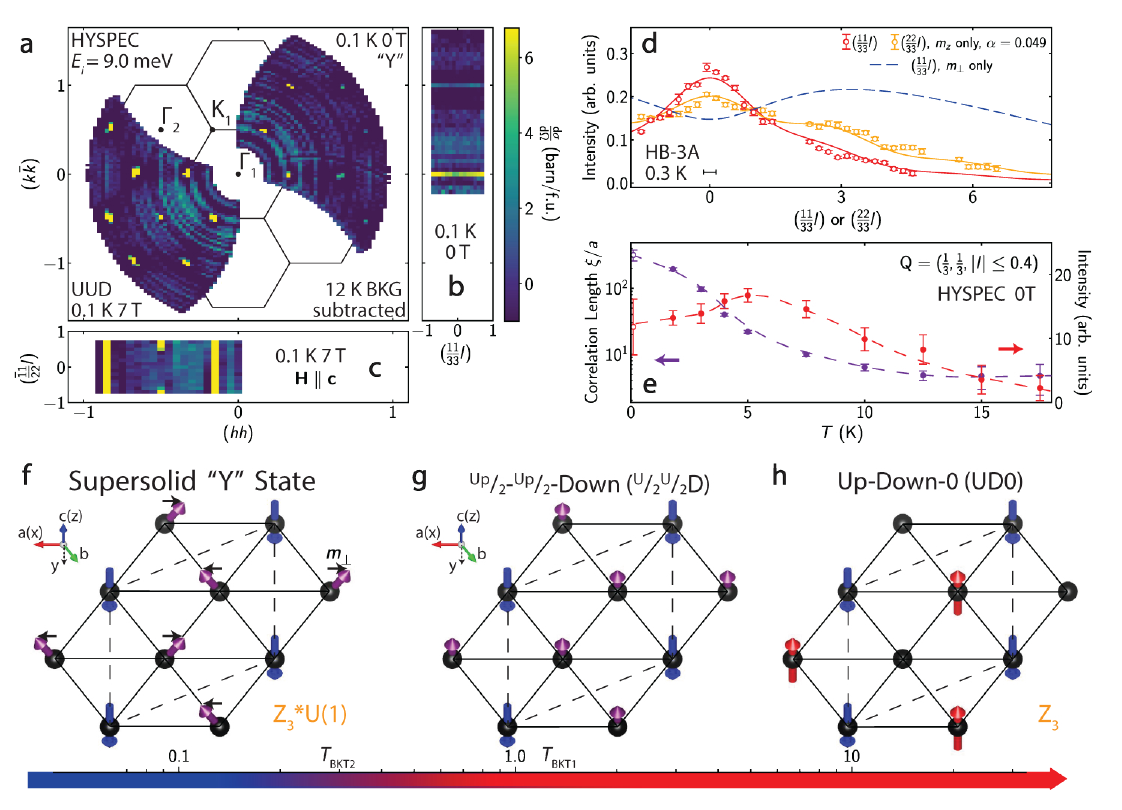}
\caption{\textbf{Elastic neutron scattering from $\rm K_2Co(SeO_3)_2$ and the inferred magnetic orders.} 
(a-c) Elastic magnetic scattering as a function of momentum, measured at $T=0.1$~K and shown after subtracting the nuclear background scattering acquired at 12 K. (a) shows scattering at 0~T (top right) and 7 T (bottom left). (b) shows 0~T data, while (c) shows 7~T data with the magnetic field applied along the easy $\bf c$-axis. 
(d) Calculated and measured $l$ dependence of magnetic neutron scattering for ${\bf Q}=(\frac{1}{3}\frac{1}{3}l)$ and $(\frac{2}{3}\frac{2}{3}l)$. 
Solid lines are fits of the ``Y'' order with $m_\perp/m_z=0.0(1)$ and $\alpha=0.049(7)$ (Eq.~\ref{Eq: elastic}).
The dashed line shows the expected intensity from a purely transverse component $m_\perp\neq 0$ ($m_z=0$ and $\alpha=0$), calculated using Eq.~\ref{Eq: elastic}.
(e) In-plane correlation length, $\xi$(in units of the lattice constant $a$), and squared staggered magnetization (intensity) as functions of temperature. The data were obtained from fits to elastic neutron scattering data acquired on HYSPEC covering an area of the $(hk0)$ plane surrounding $(\frac{1}{3}\frac{1}{3}l)$ with $|l|<0.4$. Open and closed cycles indicate data taken with $E_i=9.0$  and 3.8~meV, respectively with $|\hslash\omega|<0.5~$meV.
Dashed lines are guides to the eye. See SI for details. Error bars in (d, e) indicate the standard deviation. 
(f-h) Schematic diagrams of the candidate symmetry-breaking supersolid ``Y'', $\frac{\rm U}{2}\frac{\rm U}{2}$D, and UD0 orders discussed in the text.
} 
\label{fig:1}
\end{figure}

We start by investigating static magnetism in KCSO through elastic magnetic neutron scattering. Fig.~\ref{fig:1}a-c show background-subtracted data acquired at $T=0.1$~K in the $(hk0)$, $(hhl)$, and $(k\bar{k}l)$ reciprocal lattice planes under 0~T and 7~T magnetic fields applied along the $\bf c$-axis  (${\bf H}\parallel {\bf c}$). In zero field, despite the absence of sharp peaks in $C_{\rm p}(T)$ data, \cite{zhong2020frustrated} elastic scattering is sharply concentrated at the K points $(\frac{1}{3} \frac{1}{3})$ of the 2D Brillouin zone (Fig.~\ref{fig:1}a) and exhibits a rod-like character in the $(hhl)$ plane (Fig.~\ref{fig:1}b). This indicates quasi-2D $\sqrt{3}\times\sqrt{3}$ magnetic order. The absence of scattering at the $\Gamma$ points reveals that the dipole moment of each layer vanishes within the correlation volume defined by the $(\frac{1}{3} \frac{1}{3})$ rod. In a 7~T field, the intensity of the $(\frac{1}{3} \frac{1}{3})$ rod scattering is enhanced, and 3D Bragg peaks develop at the $\Gamma$ points (Fig.~\ref{fig:1}a,c). This indicates Up-Up-Down (UUD) type ferrimagnetic order, in which the triangular lattice planes share a uniform magnetization but otherwise remain uncorrelated with each other.

We probe the anisotropy of the magnetic order at zero field by measuring the intensity distribution of magnetic neutron scattering versus wave vector transfer $Q_z={\bf Q}\cdot\hat{\bf c}=\ell c^*$ along the rod. Fig.~\ref{fig:1}d shows the intensity decreases monotonically with $\ell$ for ${\bf Q}=(\frac{1}{3}\frac{1}{3}\ell)$ and $(\frac{2}{3}\frac{2}{3}\ell)$, with a gentle superimposed modulation that signals weak inter-layer correlations. Since the polarization factor in the magnetic neutron scattering cross section extinguishes magnetic scattering for moment ${\bf m}\parallel{\bf Q}$, the observed intensity distribution indicates the quasi-2D spin order is predominantly polarized along $\bf c$. For comparison, an in-plane spin configuration would produce the intensity distribution shown as a dashed line in Fig.~\ref{fig:1}d.

Simultaneous fits of the $(\frac{1}{3}\frac{1}{3}l)$ and $(\frac{2}{3}\frac{2}{3}l)$ data at 0.3~K yield the solid lines in Fig.~\ref{fig:1}d with weak inter-plane correlations given by $\alpha=0.049(7)$ and $m_\perp/m_z= 0.0(1)$ (Eq.~\ref{Eq: elastic}). For comparison, a theoretical calculation for $J_z/J_\perp=8$ at $T=0$ using DMRG \cite{jiang2009supersolid} yields $m_\perp/m_z= 0.27(g_\perp/g_z)=0.24$, where  $g$-factor anisotropy was obtained from high-$T$ susceptibility data.\cite{zhong2020frustrated} For KCSO where $J_z/J_\perp=14$, a smaller value of $m_\perp$ is anticipated, which may explain the lack of experimental evidence for a transverse staggered magnetization. Alternatively, it could be that $m_\perp$ only develops for $T<0.3$~K.

Since $m_\perp$ is indistinguishable from zero at 0.3~K, we can describe the spin structure in terms of Fig.~\ref{fig:1}g or Fig.~\ref{fig:1}h, which are indistinguishable based on the data in Fig.~\ref{fig:1}d. 
The inferred $z$-oriented moments on the three sites of the $\sqrt{3}\times\sqrt{3}$ unit cell are $m^{(g)}_z(1,-\frac{1}{2},-\frac{1}{2})$ with $m^{(g)}_z=3.1(3)~\mu_\textup{B}$ or $m^{(h)}_z(1,-1,0)$ with $m^{(h)}_z=2.7(3)~\mu_\textup{B}$, respectively. 
For both structures we obtain the root mean squared (RMS) ordered moment in the cell, $\sqrt{\overline{\langle m\rangle^2}}=\frac{1}{\sqrt{2}}m^{(g)}_z=\sqrt{\frac{2}{3}}m^{(h)}_z=2.2~\mu_\textup{B}$. Compared to the saturation magnetization for Co$^{2+}$ of $3.90~\mu_\textup{B}$, this corresponds to a 44(5)\% reduction in the RMS ordered moment and a strongly quantum fluctuating state. Beyond these quantitative measures, having sharp rods of scattering at ${\bf Q}_{\rm 2D}=(\frac{1}{3}\frac{1}{3})$ but not at the $\Gamma$ points (Fig.~\ref{fig:1}a,b) and no diffuse elastic scattering for $k_BT\ll J_z$ indicates a quantum fluctuating state.

Let us now examine how this state develops from the paramagnetic phase upon cooling. The $\bf Q$-integrated intensity along $(\frac{1}{3}\frac{1}{3}\ell)$ for $|\ell|\leq0.4$ -- a measure of the quasi-static staggered magnetization squared $m_z(T)^2$ -- and the in-plane magnetic correlation length, $\xi(T)$, are shown as functions of temperature in zero field in Fig.~\ref{fig:1}e (see also Supplementary Information (SI)). Upon cooling from 15~K to 5~K, both quantities increase. However, as $\xi(T)$ increases precipitously for $T<5$~K, the integrated intensity decreases slightly. This may indicate enhanced quantum fluctuations or enhanced inter-plane spin correlations that shift intensity from $l=0$ to larger $|l|$ that fall outside the $|\ell|\leq0.4$ integration range that the experiment probes.

\section*{Phase Diagram}

\begin{figure}[ht]
\centering
\includegraphics[width=1\textwidth]{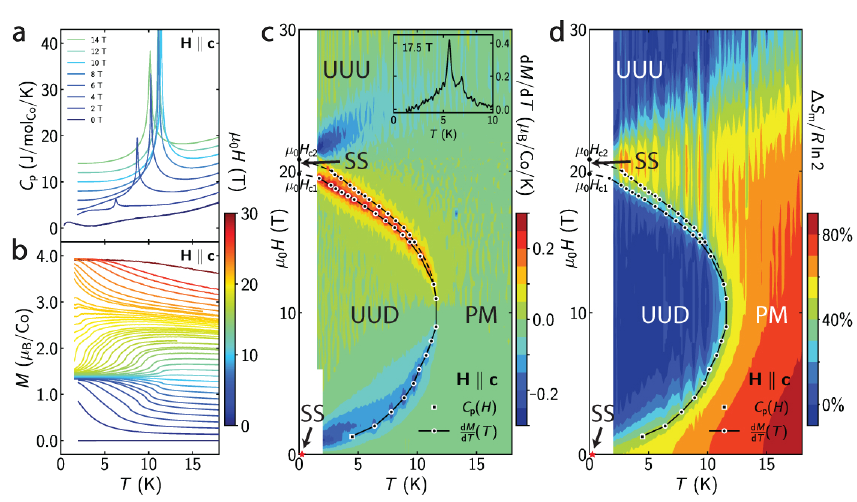}
\caption{\textbf{Temperature dependence of magnetization and specific heat capacity, and phase diagram for $\rm K_2Co(SeO_3)_2$.}
(a) Specific heat capacity as a function of temperature, with curves systematically shifted in proportion to the applied field. The zero-field specific heat capacity is reproduced with permission.\cite{zhong2020frustrated} 
(b) Magnetization versus temperature for $\bf c$-axis-oriented DC fields up to 30~T. 
(c) Interpolated color contour plot of differential susceptibility, $\textup{d}M/\textup{d}T$, versus magnetic field and temperature for ${\bf H}\parallel {\bf c}$ inferred from the data in (b).
The labels UUU, UUD, SS, and PM represent the Up-Up-Up (field-polarized), Up-Up-Down, supersolid, and paramagnetic phases, respectively. The inset shows $\textup{d}M/\textup{d}T$ as a function of temperature in a 17.5~T field. The data point on the UUD phase boundary at the lowest temperature was determined from a peak in an isothermal measurement of $C_{\rm p}(H)$.
(d) Contour plot of the magnetic entropy change, $\Delta S_m(T,H)$, normalized by the total entropy $R\ln 2$. The map was constructed by first calculating the isothermal change in entropy, $\Delta S_m(T,H)$, from $\textup{d}M(H)/\textup{d}T$ using a Maxwell relation.\cite{amaral2010estimating} The data were combined with $\Delta S_m(T,H=0)$ inferred from zero field specific heat capacity data to obtain $\Delta S_m(T,H)/R\ln 2$  (see SI).
}
\label{fig:2}
\end{figure}

We now explore the $H-T$ phase diagram for magnetic field applied along the ${\bf c}$-axis. Fig.~\ref{fig:2}a shows the heat capacity, $C_{\rm p}(T)$, for  $\mu_0H=14$~T. While the zero-field data exhibit a broad peak centered around 1.0~K, at higher fields, sharp peaks indicating a second-order phase transition emerge. These become more pronounced and shift to higher temperatures with increasing field, eventually reaching a maximum transition temperature of 11.4~K for $\mu_0 H=10$~T. Neutron diffraction (Fig.~\ref{fig:1}a,c) indicates that for fields above 2~T, this phase boundary separates the $\sqrt{3}\times \sqrt{3}$ periodic UUD and paramagnetic phases. The phase boundary is, thus, associated with breaking $Z_3$ sublattice symmetry as a D sublattice is spontaneously selected from three. For both classical \cite{noh1992phase} and quantum Ising models on a triangular lattice,\cite{yamamoto2014quantum, sellmann2015phase} this transition is predicted to be in the 2D three-state Potts universality class. Fitting the critical regime to $C_{\rm p}(T)\propto 1/(T-T_c(H))^{\alpha}$ yields $\alpha=0.318(3)$ (details in SI), which is consistent with the exact theoretical value of $\alpha = \frac{1}{3}$.\cite{wu1982potts}

Fig.~\ref{fig:2}b and \ref{fig:2}c show the uniform magnetization, $M(T)$, and differential susceptibility, d$M$/d$T$, for DC fields up to 30~T. Upon cooling in fields between 2~T and 17~T, the magnetization approaches 1/3 of the 3.90~$\mu_{\rm B}$/Co saturation value, which also indicates the UUD state. For fields beyond 22~T, the magnetization approaches saturation. The resulting $H-T$ phase diagram derived from d$M$/d$T$ (Fig.~\ref{fig:2}c) reveals a pronounced UUD phase, consistent with the classical Ising model ($J_\perp=0$) on a triangular lattice.\cite{metcalf1973phase, schick1976antiferromagnetict} In the boson representation, this corresponds to a honeycomb lattice of bosons filling 2/3 of the triangular lattice sites. Direct comparison between the measured and calculated phase diagrams yields $J_z \approx 3.0$ meV.

Using a Maxwell relation (see Methods), we obtain the field and temperature dependence of the magnetic entropy $\Delta S_m(T,H)/R\ln 2$ from the magnetization data as shown in Fig.~\ref{fig:2}d. The blue regions, where $\Delta S_m\approx 0$, distinguish gapped long-range ordered phases. Conversely, the regions where $\Delta S_m$ remains large to low temperatures indicate emergent gapless phases. For fields between 17~T and 21~T, an additional phase emerges, evidenced by a double-peak structure in the d$M$/d$T$ data versus temperature (Fig.~\ref{fig:2}c inset). This observation aligns with theoretical predictions of a high-field supersolid phase from cluster mean-field and DMRG studies of Eq.~\ref{Eq: Hxxz} with $J_{\perp} < J_{z}$.\cite{yamamoto2014quantum, sellmann2015phase} Comparing the $T \rightarrow 0$ phase boundaries ($\mu_0 H_{\rm c1}$ and $\mu_0 H_{\rm c2}$ in Fig.~\ref{fig:2}c,d) with these numerical studies \cite{yamamoto2014quantum, sellmann2015phase} indicates $J_{\perp}\approx \red{0.23}$~meV. Occurring near full magnetization, which is near full occupancy in the boson language, this high-field supersolid phase is also predicted to break both $Z_3$ and $U(1)$ symmetries. 
\red{The lower boundary of the high-field supersolid phase is predicted to be first-order for spin-$\frac{1}{2}$ quantum Ising magnets ($J_\perp < 0.4 J_z$) in numerical work.\cite{sellmann2015phase, yamamoto2014quantum} This is consistent with the steep jump observed in the $M(T)$ (Fig.~\ref{fig:2}b), and is one of the key features that distinguish KCSO from ${\rm Na_2BaCo(PO_4)_2}$.\cite{sheng2022two,gao2022spin}} 
From Fig.~\ref{fig:2}d, we estimate the change in entropy across the phase transition from the UUD phase to the putative supersolid $\Delta S_m= 0.4(1) R\ln 2$ from which the Clausius-Clapeyron relation yields the slope of the phase boundary $\mu_0$d$H_{\rm c}/$d$T_{\rm c}=-\Delta S_m/\Delta M=-0.5(1)$~T/K, which is consistent with the slope of $-0.51(1)$~T/K inferred from Fig.~\ref{fig:2}c.

\begin{figure}[ht]
\centering
\includegraphics[width=0.75\textwidth]{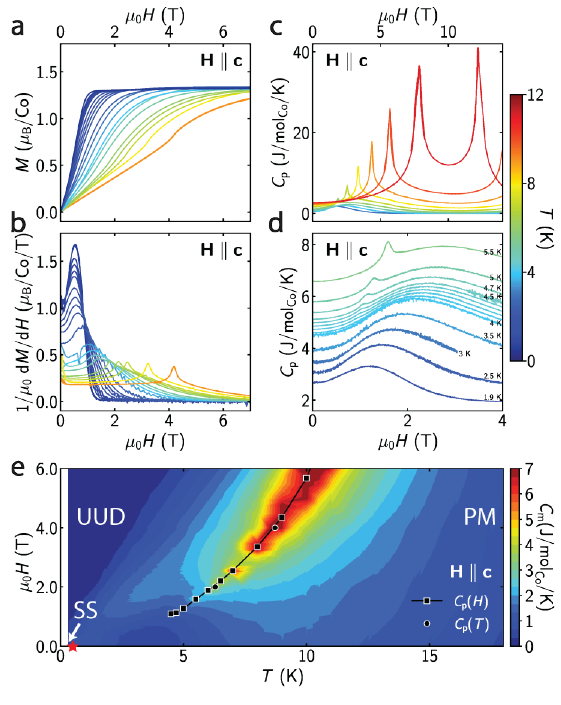}
\caption{\textbf{Magnetic field dependence of magnetization and specific heat capacity, and phase diagram for $\rm K_2Co(SeO_3)_2$.}
(a) Magnetization versus $\bf c$-axis-oriented field at various temperatures down to $T=0.5$~K. 
(b) Differential susceptibility $\textup{d}M/\textup{d}H$ versus field.
(c) Specific heat capacity versus field up to $\mu_0 H=14$~T at various temperatures.
(d) Low-field specific heat capacity as a function of field for temperatures near $T=4.5$~K.
Data are systematically shifted in proportion to temperature to show the onset of a sharp transition.
(e) Contour plot of magnetic specific heat capacity $C_{\rm m}$ versus field and temperature. The field-independent lattice contribution to the specific heat capacity was fitted by the Debye model and subtracted from $C_{\rm p}$. The second-order phase transition defined by peak positions in $C_{\rm p}$ versus temperatures (Fig.~\ref{fig:2}a) and field (Fig.~\ref{fig:3}c, d) appears to terminate at ($T$, $\mu_0H$) = (4.5~K, 1.1~T).
Data in the temperature window from 0.3~K to 1.9~K are reproduced with permission.\cite{zhong2020frustrated}}
\label{fig:3}
\end{figure}

To explore the low-field and low-temperature region of the phase diagram, we examine the magnetic field dependence of $M$, $\textup{d}M/\textup{d}H$, and $C_{\rm p}$ in Fig.~\ref{fig:3}. The $H\rightarrow 0$ differential susceptibility $\textup{d}M/\textup{d}H$ continuously increases upon cooling to the lowest temperatures accessed (0.3~K), indicating gapless magnetic excitations at low fields (Fig.~\ref{fig:3}a,b). In fact, at the lowest $H$ and $T$, $\textup{d}M/\textup{d}H$ increases with field, leading to a peak at an apparent crossover field that approaches $0.51(3)$~T at low $T$. Considering that $Z_3$ symmetry is broken in the UUD phase, the fact that this phase can be accessed through a crossover at the lowest temperature indicates that $Z_3$ symmetry is effectively broken in zero field at these temperatures. This is consistent with the long correlation length $\xi$ of the UUD phase (Fig.~\ref{fig:1}e). Fig.~\ref{fig:3}c,d show the apparent termination of the Potts transition at ($T$, $\mu_0 H$) = (4.5~K, 1.1~T), which may represent the vanishing of entropy associated with the transition rather than a critical endpoint. This would be consistent with the theoretical predictions of successive Berezinskii–Kosterlitz–Thouless (BKT) transitions associated with $Z_3$ symmetry breaking \cite{miyashita1985phase, miyashita1986magnetic, sheng1992ordering, ulaga2024finite} extending from this point to zero field (Fig.~\ref{fig:3}e). 

For temperatures above 5~K, where the crossover peak in $\textup{d}M/\textup{d}H$ is absent, $M$ increases linearly with $H$ (Fig.~\ref{fig:3}a) at low fields, and the transition to the UUD phase is marked by a sharp peak (Fig.~\ref{fig:3}b), consistent with the Potts transition. With increasing temperature, this Potts transition peak gains strength and shifts to higher fields (Fig.~\ref{fig:3}c). At 10~K, the re-entrant nature of the UUD phase is apparent, with a second sharp peak marking the upper boundary of the UUD phase within the accessible 14~T field range. Note that the putative high-field supersolid phase, identified from d$M$/d$T$ (Fig.~\ref{fig:2}c), is not yet apparent in this field regime.

\section*{Coherent Spin-Waves in the UUD Phase}

\begin{figure}[ht]
\centering
\includegraphics[width=0.75\textwidth]{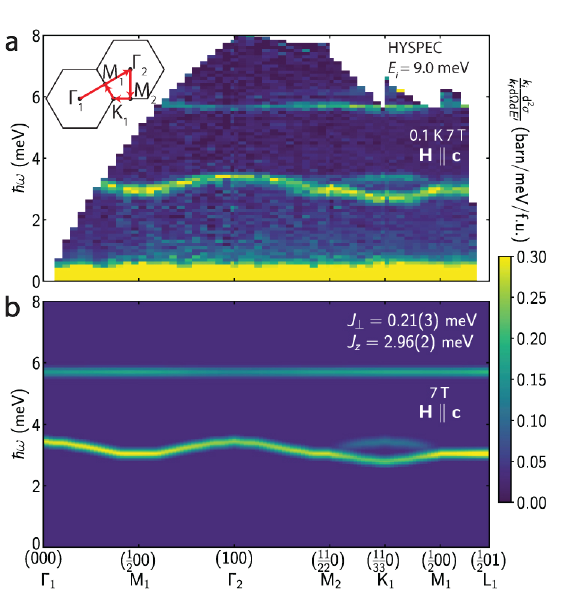}
\caption{\textbf{Magnetic neutron scattering from coherent spin waves in the UUD phase of $\rm K_2Co(SeO_3)_2$.} 
(a) The (${\bf Q},\omega$)-dependence of the magnetic neutron scattering cross section along high-symmetry directions in a 7~T magnetic field applied along the $\bf c$-axis at $T=0.1$~K.
The path through the Brillouin zone is illustrated in the inset.
Data are averaged along the $l$ direction, except for the M$_1-$L$_1$ cut along $l$.
The in-plane ${\bf Q}$ integration window is $\pm0.15$~\AA$^{-1}$ perpendicular to the trajectory.
(b) The neutron scattering cross section for $\rm K_2Co(SeO_3)_2$ in a 7 T field calculated using linear spin-wave theory, as implemented in SpinW.\cite{toth2015linear} 
The parameters $J_z = 2.96(2)$~meV and $J_{\perp} = 0.21(3)$~meV in Eqn.~\ref{Eq: Hxxz} were determined by performing a pixel-to-pixel fit of this model to the measured spectrum in panel (a).
}
\label{fig:4}
\end{figure}

A direct comparison between experiments and theory requires accurate knowledge of the underlying spin Hamiltonian. We obtain this through measurements of inelastic magnetic neutron scattering in the UUD phase at 7~T, where the magnetic scattering cross section can be calculated from the spin Hamiltonian using linear spin-wave theory. Displayed in Fig.~\ref{fig:4}, the scattering data reveal three coherent spin-wave modes, consistent with the long-range ordered $\sqrt{3}\times\sqrt{3}$ UUD state. The absence of measurable dispersion along the $\bf c$-axis (${\rm M}_1\rightarrow {\rm L}_1$ in Fig.~\ref{fig:4}a), along with the rod-like nature of the magnetic diffraction in Fig.~\ref{fig:1}c, provides clear evidence of a quasi-2D spin system.

We attribute the highest energy excitation in Fig.~\ref{fig:4}a to the flipping of a D spin, which is surrounded by $z=6$ U spins in the UUD phase. While spin-wave propagation is driven by $J_\perp$ (Eq.~\ref{Eq: Hxxz}) through the reversal of an anti-parallel spin pair, no such spin-flip process is possible after flipping an isolated D spin, which explains the dispersionless nature of the minority spin-flip excitation. The energy cost of this spin flip is approximately $2zJ_zS^2-2g_zS\mu_{\rm B}\mu_0H\approx5.84$ meV, in good agreement with the experiment (Fig.~\ref{fig:4}a). Here, $J_z\approx3.0$ meV and $g_z=7.8$ were inferred from the $H-T$ phase diagram and the saturation magnetization, respectively. The two lower-energy dispersive modes in the 7 T data are each associated with flipping one of the two U spins in the unit cell. Since U spins are surrounded by an equal number of U and D spins, the $\mathbf{Q}$-averaged energy for these modes is simply $2g_zS\mu_{\rm B}\mu_0H=3.16$~meV. Dispersion arises because the newly created D spin can move to adjacent U sites via the transverse exchange term, and the bandwidth of these lower branches in Fig.~\ref{fig:4}a provides an estimate of $3J_\perp\approx\red{0.69}$~meV.

The assignment of the upper mode to $\Delta S_z=+1$ (${\rm D}\rightarrow{\rm U}$) and the lower modes to $\Delta S_z=-1$ (${\rm U}\rightarrow{\rm D}$) is consistent with the observation that the bands move towards each other with increasing field (see Fig. S14 and S15). 
\red{As the applied field increases, the energy of the spin-flip excitation (${\rm D}\rightarrow{\rm U}$) decreases and eventually softens to zero, leading to the condensation of the upper, nearly dispersionless spin-wave mode at $\mu_0 H_{\rm c1}^{\rm calc}=6J_zS/g_z\mu_\textup{B}\approx19.9$~T. At still higher fields, the ferromagnetic spin wave in the fully polarized phase condenses at $\mu_0 H_{\rm c2}^{\rm calc}=3(2J_z+J_\perp)S/g_z\mu_\textup{B}\approx20.7$~T.\cite{gao2022spin} The separation between the two critical fields $\mu_0 H_{\rm c1}^{\rm calc}$ and $\mu_0 H_{\rm c2}^{\rm calc}$ accounts for the experimentally observed intermediate phase between the UUD and field-polarized regimes (Fig.~\ref{fig:2}c,d). Numerical studies of the spin-$\frac{1}{2}$ XXZ model on the triangular lattice\cite{yamamoto2014quantum, sellmann2015phase} identify this intermediate region as a high-field supersolid phase. While the upper critical field $\mu_0 H_{\rm c2}^{\rm calc}$ is exact at $T=0$, the lower boundary $\mu_0 H_{\rm c1}^{\rm calc}$ is predicted to become a first-order transition in quantum Ising magnets with $J_\perp < 0.4J_z$.}

To extract the most accurate values for the exchange constants, we performed a pixel-to-pixel least squares fit of linear spin-wave theory, as implemented in SpinW,\cite{toth2015linear} to the data in Fig.~\ref{fig:4}a. The calculated spin-wave cross section was convoluted with a Gaussian energy resolution with a full width at half maximum of 0.2~meV. Given the weak dispersion, effects of the finite instrumental $\mathbf{Q}$-resolution are negligible. As shown in Fig.~\ref{fig:4}b, an excellent account of the data is achieved with $J_z = 2.96(2)$ meV and $J_{\perp} = 0.21(3)$ meV. By including the next-nearest-neighbor interactions in the spin-wave calculation and comparing the resulting scattering cross section to the data, we obtained a stringent constraint on such interactions: $J_z^{(2)} = 0.02(4)$ meV and $J_{\perp}^{(2)} = 0.00(5)$ meV. The dominance of nearest-neighbor interactions is essential for exploring supersolid phases, as longer-range interactions can stabilize alternative ground states.\cite{collins1997review} The sharp spin-wave modes in Fig.~\ref{fig:4} also highlight the high quality of our multi-crystal sample and allow us to focus on intrinsic, as opposed to disorder-based, interpretations of the low-field continuum scattering.

\section*{Magnetic Excitations in Zero Field}

\begin{figure}[ht]
\centering
\includegraphics[width=\textwidth]{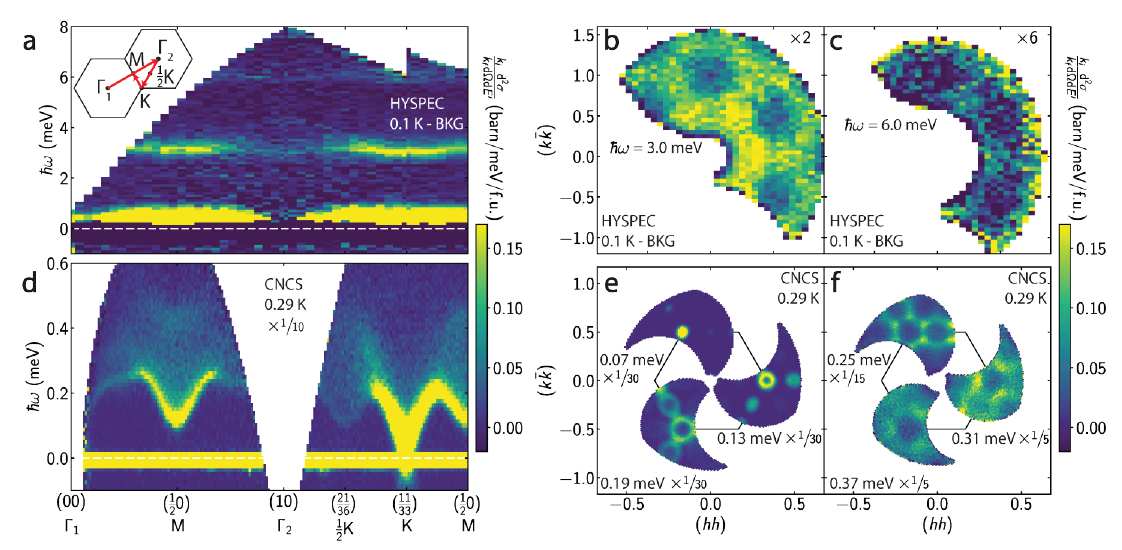}
\caption{\textbf{Zero field magnetic excitations in $\rm K_2Co(SeO_3)_2$ probed by magnetic neutron scattering.}
(a) The (${\bf Q},\omega$)-dependence of the magnetic neutron scattering cross section along high-symmetry directions in zero field at $T=0.1$~K acquired with $E_i$ = 9.0~meV neutrons. To obtain the zero-field magnetic scattering, two different backgrounds were subtracted: data from 7~T measurement for $\hslash\omega \leq 2.5$~meV, and a constant value for $\hslash\omega > 2.5$~meV, respectively. The in-plane ${\bf Q}$ integration window is $\pm0.15$~\AA$^{-1}$ perpendicular to the trajectory.
(b, c) Magnetic neutron scattering as a function of momentum in the $(hk0)$ plane for energy transfers $\hslash\omega=3$~meV and 6~meV, respectively.
A constant background was subtracted from the data.
The energy integration window is $\pm0.5$~meV.
(d) The (${\bf Q},\omega$)-dependence of the magnetic neutron scattering cross section along high-symmetry directions at $T=0.29$~K obtained with $E_i=1.0$~meV neutrons. The in-plane ${\bf Q}$ integration window is $\pm0.05$~\AA$^{-1}$ perpendicular to the trajectory.
(e, f) Low energy magnetic neutron scattering as a function of momentum in the ($hk0$) plane for six values of $\hslash\omega$.
The energy integration window is $\pm0.03$~meV.
All data shown in (a-f) have been integrated along the $l$ direction, which is justified by the quasi-2D character of the magnetic correlations.
}
\label{fig:5}
\end{figure}

Let us now examine the magnetic excitation spectrum in the low-$T$, low-$H$ limit. Though the correlation length for $\sqrt{3}\times\sqrt{3}$ quasi-static spin correlations exceeds $\xi>200~a$, where $a$ is the in-plane lattice constant, the fact that the RMS ordered moment is  reduced by as much as $44(5)$\% compared to the 7~T UUD phase indicates strong quantum fluctuations are present. Fig.~\ref{fig:5}a shows the energy and momentum dependence of these fluctuations at 0.1~K. The data reveal three distinct bands of magnetic scattering near 0~meV, 3~meV, and 6~meV. These bands exhibit some ${\bf Q}$-dependent structure (Fig.~\ref{fig:5}b,c) and are broader than the instrumental energy resolution. A fourth band at 8.9(1)~meV is inferred from THz spectroscopy (see Fig.~S15 and S16 in SI).

To understand these bands, we first consider the classical Ising model without transverse exchange ($J_\perp=0$). In this limit, magnetic excitations are individual spin-flips within the exchange field generated by the six nearest neighbors. The cost of a spin-flip is $\hslash\omega_n=J_z\Delta S^z \langle \sum_i S_i^z\rangle=J_z\langle S_{\rm tot}^z\rangle =n\times J_z$, where $n=0,1,2,3$ correspond to the four quantized values of the Ising exchange field.\cite{allan1968temperature, muttalib2024inelastic} Since a long-range ordered UUD state would only exhibits excitations corresponding to $n=0$ (for U sites) and $n=3$ (for D sites), our observation of four bands of excitations in the low-$T$ and low-$H$ limit indicates that in KCSO quantum fluctuations driven by $J_\perp$ ensure that all four possible values of the exchange-field occur (Fig.~\ref{fig:5}a). The broadening of the bands and the near-neighbor antiferromagnetic correlations indicated by the suppression of intensity at $\Gamma$-points (Fig.~\ref{fig:5}b,c) are also a result of $J_\perp$.

\begin{figure}[ht]
\centering
\includegraphics[width=1\textwidth]{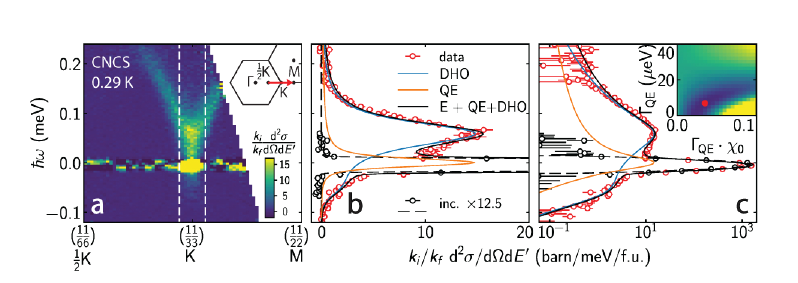}
\caption{\red{\textbf{Zero field magnetic excitations around the K point in $\rm K_2Co(SeO_3)_2$.}
(a) The (${\bf Q},\omega$)-dependence of the magnetic neutron scattering cross section along $(hh)$ at $T=0.29$~K measured with $E_i=0.72$~meV. The in-plane ${\bf Q}$ integration window is $\pm0.05$~\AA$^{-1}$ perpendicular to the trajectory. Incoherent elastic scattering has been subtracted.
(b, c) Linear and logarithmic plots of constant-$\bf Q$ cut at the K point. The red data points were obtained by averaging the data in (a) over the  the $(hh)$ range indicated by the white dashed lines. The solid black curves represent the optimal fit including a gapless and a gapped mode based on convoluting Eq.~\ref{Eq: Sqw} with the experimental energy resolution. The black data points and dashed lines indicate background subtracted incoherent elastic scattering scaled to match the integrated intensity of the magnetic Bragg peak. Therefore, the scattering beyond the dashed lines is predominantly magnetic. The inset of (c) shows $\chi^2$, quantifying the fit quality, as a function of $\Gamma_\textup{QE}$ and $\Gamma_\textup{QE}\cdot\chi_0$, which is a measure of the integrated intensity of the quasi-elastic component. The red dot indicates the minimal value for $\chi^2$. All data shown have been integrated along the $l$ direction.
}}
\label{fig:6}
\end{figure}

Focusing on the lowest energy band of scattering at $T=0.29$~K, the higher-resolution data in Fig.~\ref{fig:5}d-f show that it consists of a diffusive continuum with quasi-particle-like accumulation of intensity along its lower edge, which is consistent with a very recent study \cite{zhu2024continuum}. The strongest intensity and the lowest energy modes are found at the ordering wave vector, K. Consistent with the magnetization (Fig.~\ref{fig:3}a) and specific heat capacity data (Fig.~\ref{fig:3}e), the spectrum is gapless at K but also includes a finite-energy peak (Fig.~\ref{fig:6}). 
\red{The gapless and gapped excitations can be modeled as over- and under-damped harmonic oscillators, respectively, with the following dynamic correlation function
\begin{equation}
S(\textbf{Q},\omega) = (1+n(\omega))\frac{1}{\hslash}\bigg[\frac{\chi_0 \Gamma_{\rm QE}\omega}{\Gamma_{\rm QE}^2+\omega^2} + \frac{A}{\pi}\left(\frac{\Gamma_{\rm DHO}}{(\omega-\omega_0)^2+ \Gamma_{\rm DHO}^2}-\frac{\Gamma_{\rm DHO}}{(\omega+\omega_0)^2 + \Gamma_{\rm DHO}^2}\right)\bigg].
\label{Eq: Sqw}
\end{equation}
Here $n(\omega)=(\exp{(\hslash\omega/k_BT)}-1)^{-1}$ is the Bose-Einstein population factor. The optimal fit to the constant-$\mathbf{Q}$ cut (Fig.~\ref{fig:6}b,c) yields a relaxation rate of $\hslash\Gamma_{\rm QE} = 7(2)~\mu$eV for the quasi-elastic component, and a damping rate of $\Gamma_{\rm DHO} = 29(8)~\mu$eV with a gap $\hslash\omega_0 = 60(4)~\mu$eV for the finite energy mode. The plot of $\chi^2$ versus the parameters of the quasi-elastic component in the inset to Fig.~6c shows that both components are needed to describe the measured spectrum. In particular the instrumental energy resolution measured through nuclear incoherent elastic scattering scaled to the intensity of the truly elastic magnetic component (open circles in Fig.~6c) is sharper than the quasi-elastic component in the data for positive and negative $\hslash\omega$. The ratio of spectral weights in the three  components over the accessible energy range is $I_\textup{E}:I_\textup{QE}:I_\textup{DHO} = 1:0.023(6):0.071(5)$. The gap of the finite energy mode is found to collapse as the translational periodicity of the triangular lattice is restored by warming (see Fig. S13 in SI for details). These data are consistent with a supersolid phase wherein the truly elastic component $I_\textup{E}$ represents the frozen $\sqrt{3}\times\sqrt{3}$ Y-order, $I_\textup{QE}$ would be the gapless Goldstone mode associated with $U(1)$ rotational symmetry breaking,\cite{goldstone1962broken,weinberg1972approximate} and $I_\textup{DHO}$ would be a pseudo Goldstone mode resulting from the order-by-disorder mechanism \cite{rau2018pseudo} in the presence of quasi-long-range $\sqrt{3}\times\sqrt{3}$ order.\cite{gao2024double} However, other viable interpretations exist including that $I_{QE}$ could be gapless critical scattering associated with a thermal or quantum phase transition to a gapped phase. }

At progressively higher energies, Fig.~\ref{fig:5}d shows local spectral minima at M and at $\frac{1}{2}$K. Beyond the maximum in the dispersive lower-edge mode, there is clear evidence for continuum scattering extending to approximately 0.6~meV. All of these features can be appreciated in the $\bf Q$-dependence of constant-energy slices through the same data in Fig.~\ref{fig:5}e,f. Note that this entire spectrum is generated by quantum fluctuations and is wholly inaccessible to conventional spin wave theory. This includes the finite energy minima at M and $\frac{1}{2}$K, which, recalling the roton minimum in superfluid $^4$He, indicate proximate phases that might be stabilized with suitable additional interactions. We note that the spin wave data in (Fig.~\ref{fig:4}a) place strict limits on further neighbor interactions. The higher-energy continuum of excitations recalls the two-spinon scattering cross section for spin-$\frac{1}{2}$ chain materials,\cite{tennant1993unbound, lake2005quantum, mourigal2013fractional} suggesting that the supersolid phase might be viewed as a precursory quantum spin liquid.\cite{jia2024quantum}

\section*{Summary and Conclusions}

\K212 has proven to be an excellent platform for exploring the emergent properties of a foundational model in frustrated magnetism. In contrast to quantum spin ice, for which only rare-earth model systems are available, this Co$^{2+}(3d^7)$ based triangular-lattice system provides unprecedented experimental access to the rich emergent properties that can arise when quantum fluctuations lift the degeneracy of a frustrated spin manifold. Furthermore, a simple state -- the field-induced long-range UUD three-sublattice ordered phase -- is accessible to inelastic neutron scattering, allowing the parameters in the spin-Hamiltonian (Eq.~\ref{Eq: Hxxz}) to be accurately determined. In particular, the lack of dispersion of the minority spin-flip excitation provides direct experimental evidence that nearest neighbor interactions dominate. The ratio $J_z/J_\perp=14(2)$ places KCSO deeply in the Ising limit while the magnitude of $J_z$ ensures the emergent quantum dynamics is readily accessible to modern neutron scattering instrumentation.

Elastic neutron scattering in zero field reveals the gradual onset of solidity below 15~K, indicated by the quasi-2D $\sqrt{3}\times\sqrt{3}$ magnetic order associated with $Z_3$ translational symmetry breaking. The lack of scattering at the $\Gamma$ points (Fig.~\ref{fig:1}a,b) and the 44(5)\% reduced RMS ordered moment at 0.3~K reflect strong quantum fluctuations. The quantum fluctuations are also apparent in bands of magnetic excitations at quantized energies $\hslash\omega=n \times J_z$. At the critical K-point the coexistence of a gapless and a gapped mode at 0.29 K (Fig.~6) may be associated with the breaking of $U(1)$ spin rotational symmetry (superfluidity) and the lifting of an accidental XY degeneracy (solidity), respectively, as anticipated for a supersolid. In mapping the full $H-T$ phase diagram of KCSO, which includes a prominent UUD phase, we have also documented an additional low-$T$ phase near full magnetization that is consistent with theoretical predictions for a second supersolid phase.

In the course of finalizing this manuscript, consistent results focusing on the low-$H$ limit and low-energy spectra have been published. \cite{zhu2024continuum,zhu2025wannier} Following the posting of our manuscript on arXiv, theorists are now using advanced numerical methods, including DMRG and iPEPS, to simulate these experimental results. \cite{xu2025simulating, flores2025unconventional, mauri2025slow}

\section*{Methods}

\subsection*{Hard-Core Boson Representation}
The quantum magnetic order emerging in spin-$\frac{1}{2}$ systems on a triangular lattice, as described by the spin-Hamiltonian in Equation\eqref{Eq: Hxxz}, can be conveniently understood by representing the spin operators with hard-core bosons
\begin{equation}
\label{Eq:HCBSpinTransformation}
\begin{split}
S_i^z =& n_i - \frac{1}{2},\quad S_i^+ =& b_i^\dagger,\quad S_i^- = b_i,
\end{split}
\end{equation}
where $n_i = b_i^\dagger b_i$ is restricted to 0 or 1. Rewriting the Hamiltonian (Eq.\eqref{Eq: Hxxz}) in terms of the hard-core bosons yields:
\begin{equation}
\begin{split}
{\cal H} = \sum_{\langle i,j\rangle} \left[
V  \left(n_i-\frac{1}{2}\right)\left(n_j-\frac{1}{2}\right)
-t (b_i^\dagger b_j + b_j^\dagger b_i)
\right] - \mu \sum_i \left(n_i -\frac{1}{2}\right),
\end{split}
\label{Eq: Hxxz_bs}
\end{equation}
where the model parameters are $V=J_z$, $-t = \frac{1}{2}J_{\perp}$ and $\mu = g_z \mu_{\rm B} \mu_0 H_z$. In this bosonic representation,  $V>0$ signifies a repulsive interaction between nearest-neighbor bosons, $-t$ represents the hopping amplitude, and $\mu$ corresponds to the chemical potential. The rotational symmetry of the $x$- and $y$-components of the spins becomes $U(1)$ symmetry of the bosons.

We focus on two distinct types of orders: diagonal density order and off-diagonal phase order. Diagonal order corresponds to charge density waves (or equivalently, spatial modulations in $\langle S^z\rangle$) and is often referred to as solidity.
Off-diagonal order involves the spontaneous breaking of the $U(1)$ symmetry and is characteristic of superfluidity.
According to the Mermin-Wagner theorem, true long-range order associated with a continuous symmetry cannot develop at finite temperatures in a 2D system. 
In the superfluid phase that exhibits a quasi-long-range order, the correlation length is limited by thermally generated bound vortex–antivortex pairs.

The antiferromagnetic Heisenberg model on triangular lattices with easy-axis anisotropy has been suggested to host phases exhibiting both diagonal and off-diagonal long-range order, referred to as supersolids. Since $m_\perp=g_\perp \langle S_\perp \rangle$ is linked to the creation and annihilation of bosons ($b^\dagger$ and $b$), a nonzero $m_\perp$ reflects a macroscopic quantum coherence of the bosons, which is indicative of superfluidity. Similarly, a nonzero $m_z=g_z\langle S_z \rangle$ reflects the spatial modulation of boson density, which is analogous to a classical charge density wave in a solid.

\subsection*{Materials Synthesis}
$\rm K_2Co(SeO_3)_2$ single crystals were synthesized using a previously reported solid-state reaction method \cite{zhong2020frustrated} at Johns Hopkins University and Princeton University.
Neither stacking faults nor secondary phases were detected in X-ray diffraction analysis of polycrystalline samples.  
The crystals were found to crystallize in the space group $R\bar{3}m$ (No. 166) with lattice constants $a = b = 5.5049(7)$~\AA, and $c = 18.411(3)$~\AA\ at 100~K.

\subsection*{Specific Heat}
The specific heat capacity of KCSO was measured using a thermal-relaxation method in a Quantum Design Physical Properties Measurement System (PPMS) at Johns Hopkins University. A 2.1~mg plate-like sample was mounted with its primary face horizontal to align the $\bf c$-axis with the vertical field direction of the 14~T magnet.

\subsection*{Magnetization}
High-field DC magnetization data, $M(T)$, were obtained at the National High Magnetic Field Laboratory (NHMFL) in Tallahassee, Florida. The measurements used a conventional vibrating sample magnetometer (VSM) in a water-cooled resistive magnet located in Cell 8 of the DC Field Facility. The VSM was calibrated using a standard Ni sphere, and the sample was glued to a polycarbonate sample holder with GE7031 varnish. The magnetization data were normalized against DC magnetization measurements performed with a VSM in a 14~T PPMS (Quantum Design) at Johns Hopkins University.

From the Maxwell relation $(\partial S/\partial H)_T=\mu_0(\partial M/\partial T)_H$, we can derive the isothermal change in entropy $\Delta S_m(H)=\mu_0\int_0^H(\partial M/\partial T)_{H'}\textup{d}H'$.\cite{amaral2010estimating} By combining this with the change in entropy $\Delta S_m(T,H=0)$ inferred from zero-field specific heat capacity data, we obtain the full field and temperature dependence of the entropy $\Delta S_m(T,H)/R \ln 2$.

The magnetization, $M(H)$, in millisecond pulsed magnets reaching up to 60~T were measured at the NHMFL pulsed field facility at Los Alamos National Laboratory. The experiments were conducted in three configurations with magnetic fields along the $\bf c$-axis direction. To reduce the background from the spatially uniform pulsed magnetic field, the samples were placed in a radially counter-wound copper coil. An additional one turn of the coil further compensated for any residual signals. For each temperature, two measurements were taken: one with the sample inside the coil and one with it outside. The final magnetization curve was obtained by subtracting the ``sample-out'' background signal from the ``sample-in'' data. The temperature was stabilized using a $^3$He system, and a Lakeshore Cernox thermometer recorded the temperature before each field pulse. The pulsed-field data in Fig. S5  clearly display two high field maxima in $dM/dH$ indicating an additional phase near magnetization saturation. The rapid magnetization and demagnetization processes resulted in considerable sample heating and cooling, respectively, consistent with suggestions to use such materials for cryogenic refrigeration.\cite{xiang2024giant} For this reason, we chose to base the phase diagram in Fig.~\ref{fig:2} on DC magnetization data only.

\subsection*{Elastic Neutron Scattering}
Neutron diffraction experiments were conducted using the HB-3A DEMAND diffractometer \cite{cao2018demand} at the High Flux Isotope Reactor at Oak Ridge National Laboratory. 
An 8.22~mg single crystal was mounted on an oxygen-free high-thermal-conductivity (OFHC) copper holder and cooled using a $^3$He insert with a base temperature of 0.3~K.
The experiment used a four-circle mode and a beam of neutrons with a wavelength $\lambda=1.542$ \AA\ from a bent Si-220 monochromator.

Fig.~\ref{fig:1}d shows the magnetic elastic scattering versus  ${\bf Q}=(\frac{1}{3}\frac{1}{3}l)$ and $(\frac{2}{3}\frac{2}{3}l)$, measured in zero field without final energy analysis. The broad peaks versus $l$ indicate quasi-2D magnetic order with moments predominantly oriented along the $\bf c$-axis. The dashed and solid lines in Fig.~\ref{fig:1}d represent calculations based on the following expression for neutron diffraction from an anisotropic quasi-2D  magnetic structure with ($\alpha\neq0$) and without ($\alpha=0$) inter-plane correlations:
\begin{equation}
\begin{aligned}
I(\textbf{Q}) = N_{\rm M} \frac{(2\pi)^2}{A_{\rm M}} (\frac{\gamma r_0}{2})^2 |F(\textbf{Q})|^2 \times 
(1+2 \alpha\cos(\frac{2\pi l}{3})) \times 
\Big (
\underbrace{(1-{\hat Q}_z^2)|\mathbfcal{F}_z(\textbf{Q})|^2}_{\textup{solidity}} 
 + \underbrace{\frac{1}{2}(1+{\hat Q}_z^2)|\mathbfcal{F}_\perp(\textbf{Q})|^2}_{\textup{superfluidity}}
 \Big ).
\end{aligned}
\label{Eq: elastic}
\end{equation}
This expression is based on the following approximation $\langle S_i^{\alpha}(t) S_j^{\beta}(0)\rangle\approx \langle S_i^{\alpha}\rangle\langle S_j^{\beta}\rangle$ and averaging over all domains. $N_{\rm M}$ and $A_{\rm M}$ are the number and in-plane area of magnetic unit cells; $\gamma r_0 = -0.54 \cdot 10^{-12}$ cm is the magnetic scattering length; $\textbf{Q}$ is the momentum transfer; $(1+2 \alpha\cos(\frac{2\pi l}{3}))$ accounts for inter-plane correlations; and $(1-{\hat Q}_z^2)$ and $\frac{1}{2}(1+{\hat Q}_z^2)$ are domain averaged polarization factors for the two components of the magnetic structure. $F(\textbf{Q})$ is the magnetic form factor, which we approximate as the atomic form factor for Co$^{2+}$. \cite{clementi1974roothaan}

$\mathbfcal{F}_{z,\perp}(\textbf{Q})=\sum_j m^{(j)}_{z,\perp}\exp (i{\rm\bf Q}\cdot{\rm\bf d}_j)$ are the scalar magnetic structure factor for the in- and out-of-plane components of the dipole moments $m^{(j)}_{z,\perp}$ at locations ${\rm\bf d}_j$ within the $\sqrt{3}\times\sqrt{3}$ magnetic unit cell. For all allowed magnetic Bragg peaks of the structures in Fig.~\ref{fig:1}f-h we have $|\mathbfcal{F}_{z}^{(f,g)}|^2=\frac{9}{4}m_{z}^2$, $|\mathbfcal{F}_{\perp}^{(f)}|^2=3m_{\perp}^2$, and $|\mathbfcal{F}_{z}^{(h)}|^2=3m_{z}^2$. Here $m_{z,\perp}$ denote the largest parallel and perpendicular component of the ordered dipole moment, respectively.

\subsection*{Inelastic Neutron Scattering}
The neutron scattering data were collected on the HYSPEC and CNCS spectrometers at Oak Ridge National Laboratory.
For the HYSPEC experiment, single crystals with a total mass of 0.9~g were co-aligned on an aluminum mount for scattering in the $(hk0)$ reciprocal lattice plane. The sample was cooled in a dilution refrigerator with an 8~T vertical field magnet to a base temperature of 70~mK.
Measurements used incident energies $E_i=$3.8~meV and 9.0~meV with a 240~Hz chopper frequency. The sample was rotated through 60$^\circ$ or 120$^\circ$ in 2$^\circ$ steps. 
In the 0.1~K ``zero-field'' measurement, a 0.02~T field was applied to maintain the aluminum sample mount in its thermally conductive normal state. 
Data were normalized against the magnetic Bragg diffraction intensity in the plateau phase, where the sublattice magnetization is known to be $m_z=3.90~\mu_{\rm B}$. Measurements on CNCS employed a $^3$He insert in a cryostat with a base temperature of 0.29~K. $E_i=0.72$~meV and 1.0~meV were used with a 300~Hz chopper frequency in high-flux mode. The sample was rotated through 30$^\circ$ or 60$^\circ$ in 1$^\circ$ steps. The CNCS data were normalized to the HYSPEC measurements using the incoherent elastic scattering cross section.

\subsection*{THz Spectroscopy}
Time-domain terahertz spectroscopy measurements were performed in a home-built system equipped with a commercial fiber-coupled Toptica spectrometer and a 6.5 T superconducting magnet.\cite{tagay2024high} The magnetic field was applied along the {\bf c}-axis. The complex THz transmission matrix was measured in a frequency range from 0.2~THz to 2~THz.

\bibliography{main_ref}

\noindent\textbf{Acknowledgements}\\
We gratefully acknowledge valuable discussions with Gang Chen, Cristian Batista, Yuan Gao, Andreas L{\"a}uchli, Wei Li, Changle Liu,  Fr\'ed\'eric Mila, Roderich Moessner, Oleg Tchernyshyov, and Shu Zhang. Initial phases of this work were supported as part of the Institute for Quantum Matter, an Energy Frontier Research Center funded by the U.S. Department of Energy, Office of Science, Basic Energy Sciences under Award No. DE-SC0019331. Further neutron scattering work was supported by Department of Energy, Office of Science, Basic Energy Sciences under Award No. DE-SC0024469. C.B. was supported by the Gordon and Betty Moore Foundation EPIQS program under GBMF9456. Y.H. and H.C. were supported by the U.S. Department of Energy, Office of Basic Energy Sciences, Early Career Research Program Award KC0402020. J.Z. acknowledges the support of the NSF CAREER grant DMR-1848349. A portion of this work was performed at the National High Magnetic Field Laboratory, which is supported by the National Science Foundation Cooperative Agreement No. DMR-2128556*, the U.S. Department of Energy, and the State of Florida. This research used resources at the High Flux Isotope Reactor and Spallation Neutron Source, DOE Office of Science User Facilities operated by Oak Ridge National Laboratory. The beam time was allocated to HYSPEC and CNCS on proposal number ITPS-29655. The beam time was allocated to HB-3A on proposal number IPTS-31928.\\

\noindent\textbf{Author contributions}\\
T.C., R.Z., and C.B. initiated this work. A.G., X.X., and R.C. prepared the samples. T.C., Y.C., Y.H., H.C., B.W., A.P., and D.P. carried out neutron scattering experiments. A.G., E.C., M.J., and M.L. measured high-field magnetization. L.S., Z.T., and N.P.A. performed THz measurements. T.C., A.G., J.Z., L.C., and C.B. wrote the manuscript with input from all coauthors.\\

\noindent\textbf{Competing interests}\\
The authors declare no competing interests.\\

\section*{Figures}

\end{document}


\preprint{APS/123-QED}

\title{Supplementary Information for Phase Diagram and Spectroscopic Signatures of a Supersolid in the Quantum Ising Magnet \K212}

\author{Tong Chen}\email{tchen115@jhu.edu}
\author{Alireza Ghasemi}
\author{Junyi Zhang}
\author{Liyu Shi}
\author{Zhenisbek Tagay}
\author{Youzhe Chen}
\author{Lei Chen}
\author{Eun-Sang Choi}
\author{Marcelo Jaime}
\author{Minseong Lee}
\author{Yiqing Hao}
\author{Huibo Cao}
\author{Barry Winn}
\author{Andrey A. Podlesnyak}
\author{Daniel M. Pajerowski}
\author{Ruidan Zhong}\email{rzhong@sjtu.edu.cn}
\author{Xianghan Xu}
\author{N. P. Armitage}
\author{Robert Cava}
\author{Collin Broholm}\email{broholm@jhu.edu}

\date{\today}
\maketitle

\section{\label{appendix: Critical Exponents} Critical Exponents}

At finite fields, sharp anomalies in the specific heat capacity versus temperature indicate a second-order transition from the paramagnetic state to the long-range UUD phase, as shown below. To extract the associated critical exponent, we perform a least-squares fitting analysis accounting for instrumental effects that broaden the peak.

\begin{figure}[H]
    \centering
    \includegraphics[width=1\columnwidth]{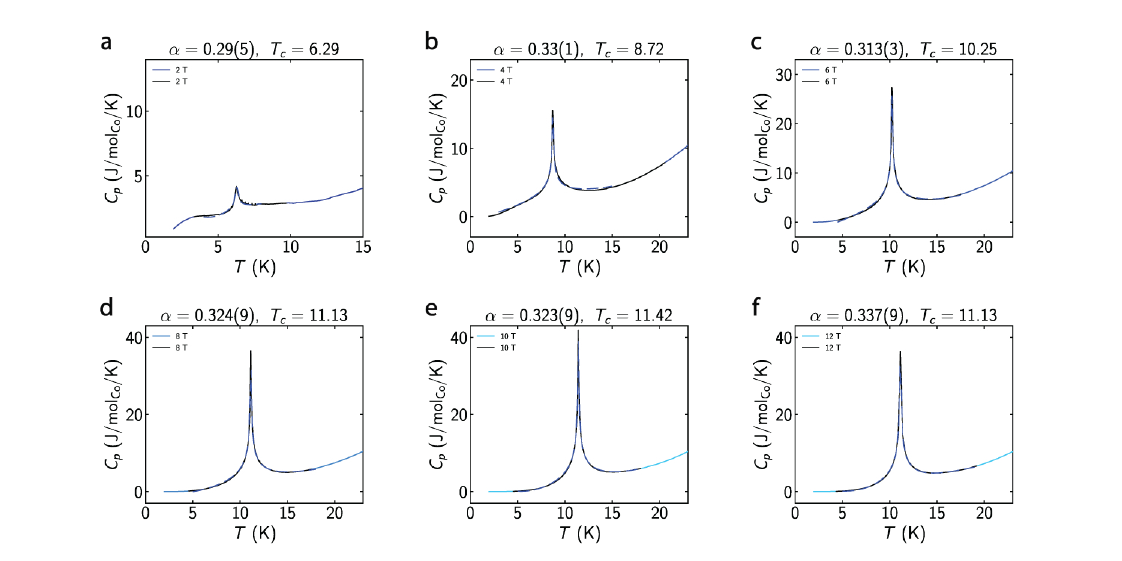}
    \caption{Representative plots of $C_{\rm p}$ versus $T$. Blue-ish lines indicate measured specific heat. Black lines indicate the regions for fitting. Blue dashed lines are fittings against Eq.~\ref{Eq: CvT}. 
    } \label{sfig:HCvsT}
\end{figure}

The Quantum Design Heat Capacity system used for these measurements employs an adiabatic relaxation technique in which each cycle starts with a heat pulse applied to the sample, followed by a period of relaxation during which the sample stage returns to thermal equilibrium with the cryogenic reservoir. After each measurement cycle, the entire temperature response of the sample platform is fitted to a model that accounts for thermal relaxation both of the sample platform to the reservoir and between the sample platform and the sample itself. Therefore, the specific heat capacity assigned to each temperature is effectively an average over a small temperature range, which broadens the critical behavior of specific heat. To account for this, we convolve the divergent critical function with a Gaussian:
\begin{equation}
\begin{aligned}
C(t) &= A_\textup{T}  |t|^{-\alpha}\\
C_p(t) = \Sigma_{i=0}^4 a_i t^i + &\int C(t-\tau) e^{-\tau^2/(2\sigma_\textup{T}^2)} \textup{d} \tau,
\label{Eq: CvT}
\end{aligned}
\end{equation}
Here, the first polynomial term accounts for the non-singular part of the specific heat, $\sigma_T\sqrt{8\ln2}$ is an empirical full-width at half maximum (FWHM)  Gaussian smearing associated with the heat pulse method, and $A_T$ is a pre-factor controlling the strength of the critical term.

\begin{figure}[H]
    \centering
    \includegraphics[width=1\columnwidth]{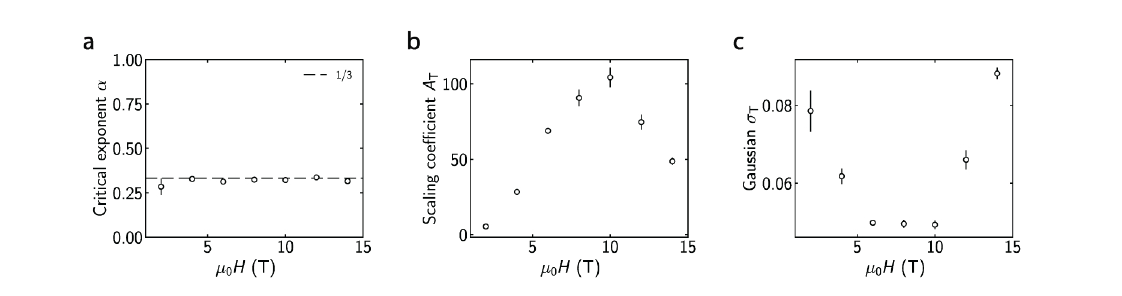}
    \caption{Field dependence of critical exponent $\alpha$, scaling coefficient $A_\textup{T}$, and standard deviation $\sigma_\textup{T}$ from the fitting. 
    } \label{sfig:HCvsTexp}
\end{figure}

Similarly, we define a convoluted critical function for the specific heat versus field:
\begin{equation}
\begin{aligned}
C(h) &= A_\textup{H}  |h|^{-\alpha'}\\
C_p(h) = \Sigma_{i=0}^4 c_i h^i + &\int C(h-h') e^{-h'^2/(2\sigma_\textup{H}^2)} \textup{d} h',
\label{Eq: CvB}
\end{aligned}
\end{equation}
where the theory of critical phenomena indicates that $\alpha'=\alpha$. $2\sqrt{2\ln2}\sigma_H$ is the empirical FWHM smearing of the field-driven transition that should be related to the thermal smearing as follows: $\sigma_B\approx\sigma_T(\mu_0\textup{d}H_c(T)/\textup{d}T)$. Here $\mu_0 H_c(T)$ is the $T$-dependent phase boundary for the UUD plateau phase. $A_H$ regulates the strength of the critical term. 

\begin{figure}[H]
    \centering
    \includegraphics[width=1\columnwidth]{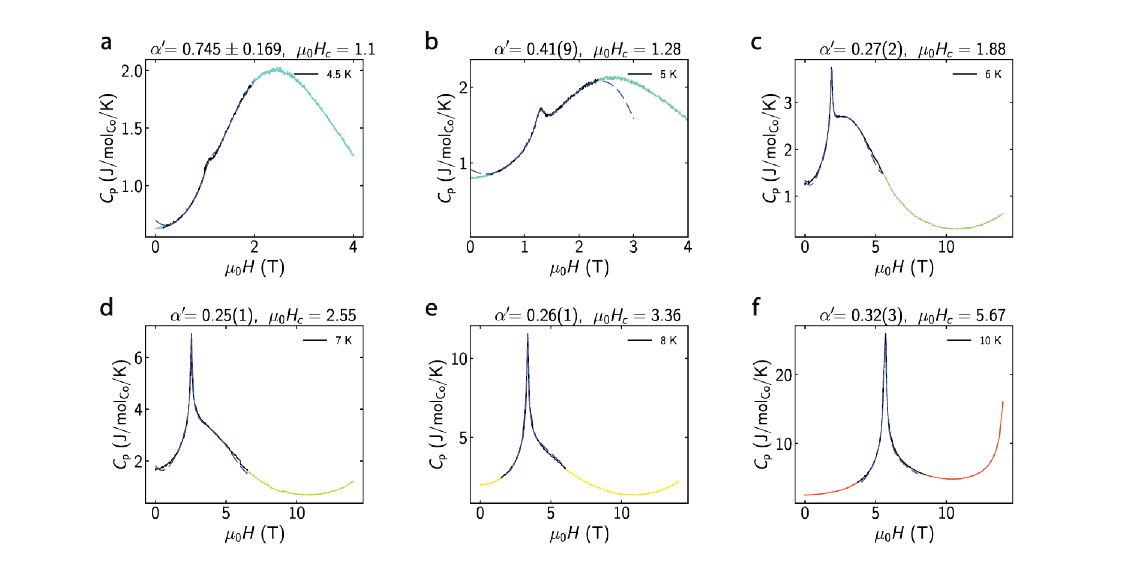}
    \caption{Representative plots of $C_p$ versus $H$. Black lines indicate the regions for fitting. Blue dashed lines are fittings against Eq.~\ref{Eq: CvB}. 
    } \label{sfig:HCvsB}
\end{figure}

For the 3-state Potts model in two dimensions, critical exponents are known to be $\alpha=1/3$, $\beta=1/9$, $\gamma=13/9$, and $\delta=14$ \cite{wu1982potts}. Our fit yields a critical exponent of $\alpha=0.318(3)$ (Fig.~\ref{sfig:HCvsTexp}) for specific heat versus temperature, which aligns with the expected value for the 2D three-state Potts universality class.

From the fits to $C_p(H)$, we obtain an average critical exponent of $\alpha'=0.27(1)$. This deviation from $\alpha= 1/3$ may be due to systematic errors associated with the heat pulse method when the phase boundary slope, $\mu_0\textup{d}H_c(T)/\textup{d}T$, is large.

\begin{figure}[H]
    \centering
    \includegraphics[width=1\columnwidth]{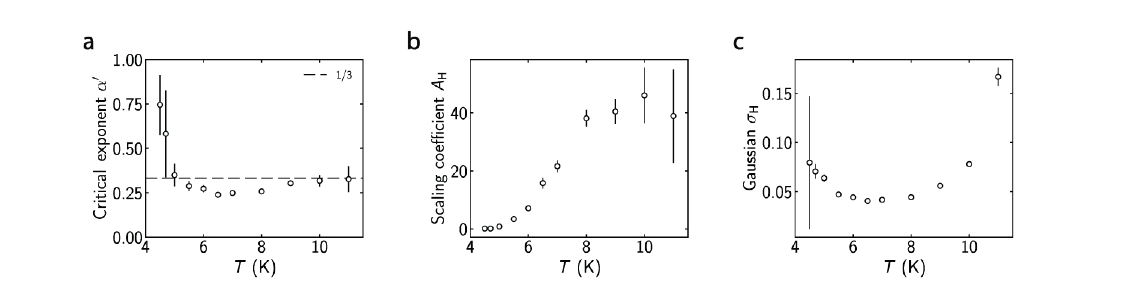}
    \caption{Temperature dependence of critical exponent $\alpha'$, scaling coefficient $A_\textup{H}$, and standard deviation $\sigma_\textup{H}$ from the fitting. 
    } \label{sfig:HCvsBexp}
\end{figure}


\section{\label{appendix: Magnetization} High-Field Magnetization}

\begin{figure}[H]
    \centering
    \includegraphics[width=1\columnwidth]{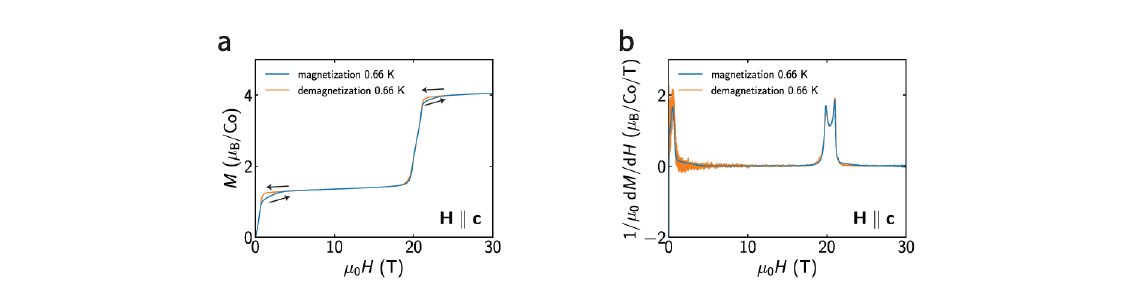}
    \caption{Representative magnetic field dependence of magnetization, $M(H)$, and differential susceptibility, $\textup{d}M/\textup{d}H$, up to 30~T measured at 0.66~K.
    } \label{sfig:MvsH}
\end{figure}

The magnetization, $M(H)$, was measured in millisecond pulsed magnets reaching up to 60~T at the NHMFL pulsed field facility at Los Alamos National Laboratory. For these measurements, the temperature was stabilized using a $^3$He system, with a Lakeshore Cernox thermometer recording the temperature before each pulse. Representative $M(H)$ and $\textup{d}M/\textup{d}H$ curves, measured under magnetic field along the ${\bf c}$-axis at a starting temperature of 0.66~K, are shown in Fig.~\ref{sfig:MvsH}. The data reveal a pronounced 1/3 magnetization plateau in $M(H)$ and a double-peak structure in $\textup{d}M/\textup{d}H$. This structure indicates the high-field supersolid phase in $\textup{d}M/\textup{d}H$ and is consistent with the $M(T)$ and $\textup{d}M/\textup{d}T$ data shown in the main text (Fig. 2b,c). However, because the rapid magnetization and demagnetization process resulted in considerable sample heating and cooling, respectively, we constructed the phase diagrams in Fig. 2 using only DC magnetization data.

\section{\label{appendix: Correlation Length} Correlation Length}

We now present the method used to determine the correlation length from elastic scattering obtained at HYSPEC. Assuming the UUD phase represents a long-range order, we fit the $(\frac{1}{3}\frac{1}{3})$ elastic peak measured with $E_i=3.8$~meV at 1.8~K to a constant background plus 2D Gaussian function of the following form:
\begin{equation}
\begin{aligned}
G({\bf q-q}_0) = \frac{A'}{2\pi \sigma_x \sigma_y} \exp\left( -\frac{1}{2}\left( (\frac{x-x_0}{\sigma_x})^2 + (\frac{y-y_0}{\sigma_y})^2\right)\right).
\label{Eq:2DGaussian}
\end{aligned}
\end{equation}
Here $x,y$ indicate wave vector transfer along the perpendicular $(xx0)$ and $(y\bar{y}0)$ directions, respectively.  $2\sqrt{2\ln2}\sigma_x$ is thus a measure of the FWHM instrumental resolution at ${\bf q}=(\frac{1}{3}\frac{1}{3}0)$ along the longitudinal ($hh0$) direction while $2\sqrt{2\ln2}\sigma_y$ is the FWHM along the transverse ($k\bar{k}0$) direction, which is mainly determined by the sample mosaic distribution.
Fits to 3.5~T and 4~T data yield $\sigma_x = 0.0046(1)$~r.l.u. and $\sigma_y=0.0132(1)$~r.l.u. The corresponding widths are $\sigma_x=0.0046\sqrt{3}a^*=0.010$~\AA$^{-1}$ and $\sigma_y=0.0132(1)a^*=0.017$~\AA$^{-1}$. 

To assess the spin correlation function of $\rm K_2Co(SeO_3)_2$ versus temperature, we analyzed 2D momentum-dependent magnetic neutron scattering data acquired with $E_i=3.8$~meV at HYSPEC. The fitting function was the convolution of a Gaussian resolution function, $G({\bf q-q}_0)$ \eqref{Eq:2DGaussian}, with a model for the intrinsic spin correlation function, $\cal{S}({\bf q})$:
\begin{equation}
    {\cal I}({\bf q})=\int {\rm d}^2{\bf q}' G({\bf q'-q}) {\cal S}({\bf q'-q}_0),
    \label{eq:conv}
\end{equation}
We used the following isotropic form for the intrinsic spin correlation function: 
\begin{equation}
{\cal S}({\bf q-q}_0)=\frac{\kappa^{2n-2}}{(|{\bf q}-{\bf q}_0|^2+\kappa^2)^n}, 
\label{eq:lorentz}
\end{equation}
\label{Eq:conv_Lorentzian}
where ${\bf q}_0=(\frac{1}{3}\frac{1}{3}0)$ and we tested both $n=1$ and $n=2$. The corresponding isotropic real space correlation function is obtained by Fourier transformation: 
\begin{eqnarray}
{\cal S}(r)&=&\int \textup{d}^2{\bf q}{\cal S}({\bf q})e^{-i{\bf q}\cdot{\bf r}}\\
&=&\int_{0}^{\infty}q\textup{d}q{\cal S}(q)\int_0^{2\pi}\textup{d}\phi e^{-iqr\cos{\phi}}\\
&=&2\pi\int_{0}^\infty q\textup{d}q{\cal S}(q)J_0(qr)\\
&\equiv&g(\kappa r),
\end{eqnarray}
Here we set aside the antiferromagnetic modulation associated with ${\bf q}_0$ and use the continuum limit, which is appropriate for examining the large-$r$ limit of the correlation function. The correlation function takes the form
\begin{equation}
g(x)=2\pi\int_0^\infty \frac{uJ_0(ux)}{(u^2+1)^n}du
\label{eq:gx}
\end{equation}
Defining a spatial correlation length as follows:
\begin{equation}
\xi=-\underset{r\rightarrow\infty}{\lim}\frac{r}{\log(g(\kappa r))}\equiv\frac{\eta(n)}{\kappa},
\end{equation}
 we obtain $\eta(1)=1.1$ and $\eta(2)=0.93$ through numerical integration of Eq.~\ref{eq:gx}. We report $1/\kappa$ as the inferred correlation length for $n=1$ and $n=2$.

\begin{figure}[H]
    \centering
    \includegraphics[width=1\columnwidth]{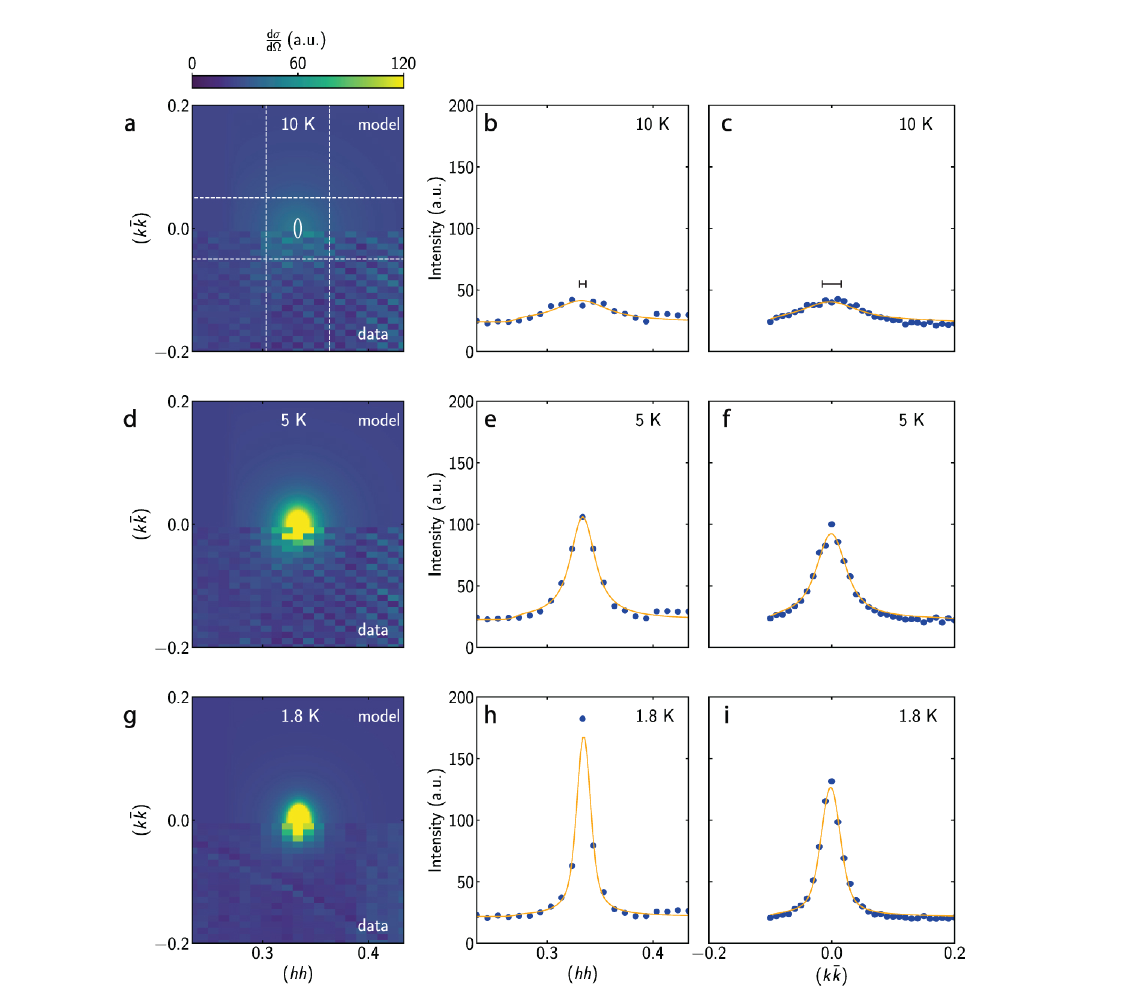}
    \caption{Representative data and fits using Eq.~\ref{eq:conv} at various temperatures. Solid and dashed lines in the two right columns correspond to $n=1$ and $n=2$, respectively in Eq.~\ref{eq:lorentz} with $n=1$ providing a marginally better fit. Dashed lines in (a) indicate the integration area for the cuts along $(hh0)$ and $(k\bar{k}0)$. The white ellipse in panel a and horizontal bars in panels b,c indicate the full width at half maximum (FWHM) of the instrumental resolution.  
    } \label{fig:sfig6}
\end{figure}

\begin{figure}[H]
    \centering
    \includegraphics[width=1\columnwidth]{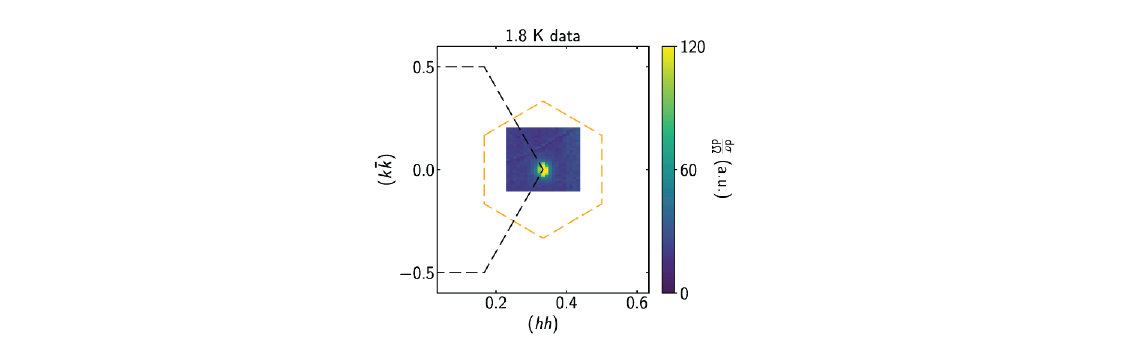}
    \caption{Diffraction data at 1.8~K. In reporting the integrated intensity in Figure 1h, we integrated the model function in Eq~\ref{eq:lorentz} over the region indicated by the dashed orange line.
    } \label{fig:sfig7}
\end{figure}

\section{\label{appendix: Inelastic Neutron Scattering} Inelastic Neutron Scattering}

Inelastic neutron scattering data for $\rm K_2Co(SeO_3)_2$ were acquired on the HYSPEC spectrometer at Oak Ridge National Laboratory. A 0.9~g co-aligned assembly of single crystals was mounted for scattering in the $(hk0)$ plane and installed in a dilution refrigerator, with a base temperature of 70~mK and an 8~T vertical-field split-coil magnet. The data were acquired with incident energies $E_i = 3.8$~meV and 9.0 meV and a 240 Hz chopper frequency. According to simulations using MANTID \cite{arnold2014mantid}, the energy resolutions at the elastic line were 0.09 meV and 0.29 meV, respectively. We used the (110) and $(2\bar{1}0)$ nuclear Bragg peaks to align the sample on the instrument.

\begin{figure}[H]
    \centering
    \includegraphics[width=0.9\columnwidth]{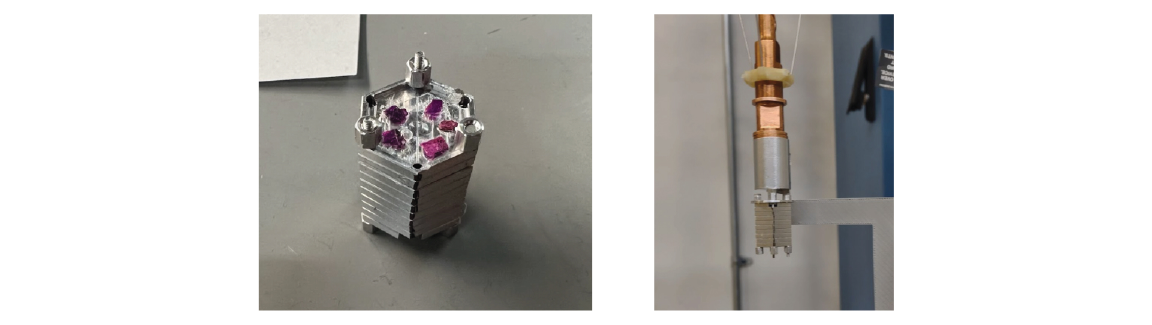}
    \caption{Photograph of the co-aligned  assembly of 0.9 g $\rm K_2Co(SeO3)_2$ crystals used for inelastic neutron scattering on HYSPEC. 
    } \label{fig:sfig8}
\end{figure}

\begin{figure}[H]
    \centering
    \includegraphics[width=1\columnwidth]{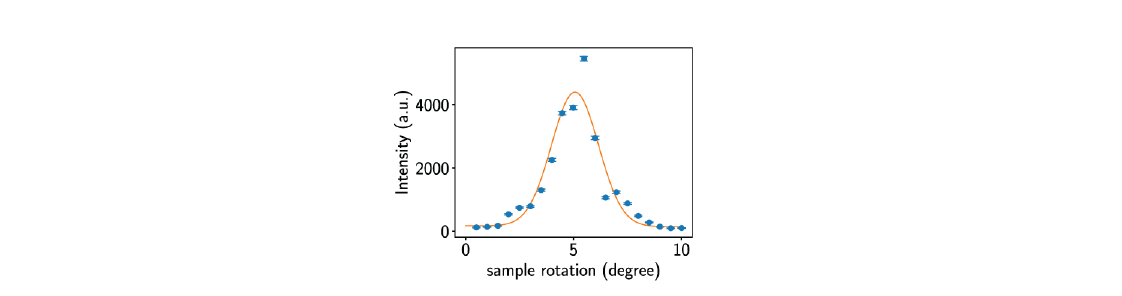}
    \caption{A rocking scan through the (2$\bar{1}$0) nuclear Bragg peak measured on HYSPEC. A Gaussian fit yields a FWHM of 2.6(2)$^\circ$.
    } \label{fig:sfig9}
\end{figure}

\begin{figure}[H]
    \centering
    \includegraphics[width=1\columnwidth]{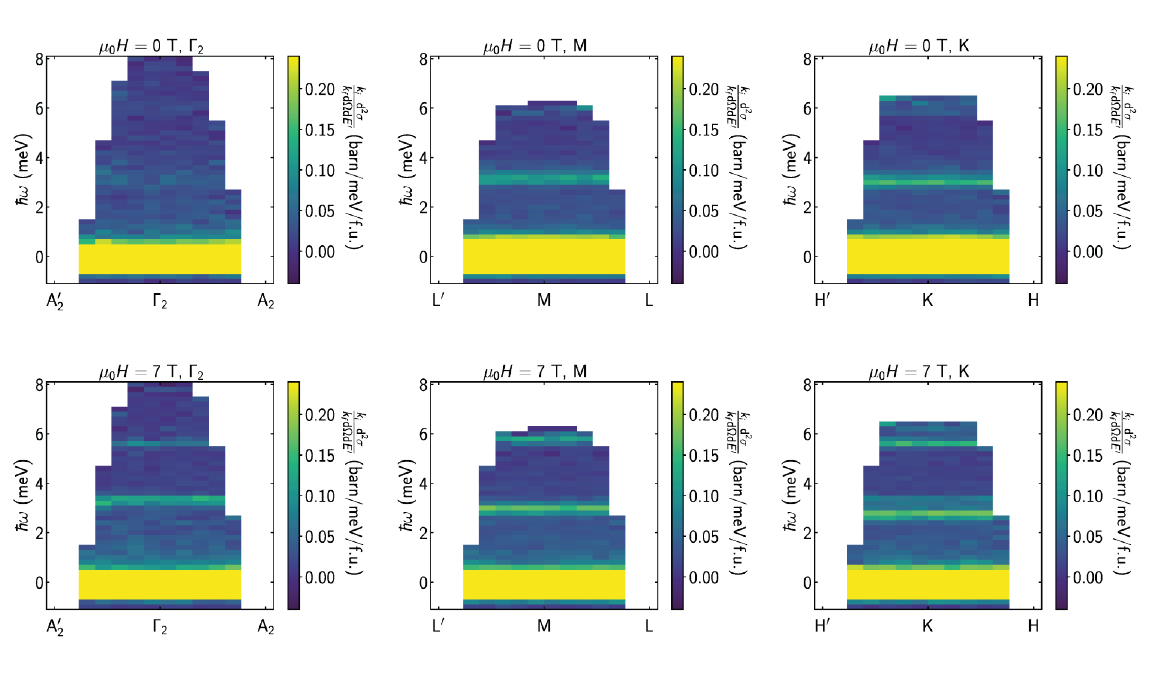}
    \caption{Inelastic neutron scattering from $\rm K_2Co(SeO3)_2$ as a function of energy transfer $\hbar\omega$ and momentum transfer along high-symmetry lines that extend along $\bf c^{\rm *}$. 
    The averaging windows are $\Delta^2\textbf{Q}=(\pm0.1,\pm0.1,0)$.
    The c-axis aligned magnetic field is 0~T and 7~T  at $T=0.1$~K. The incident neutron was $E_i=9.0 $~meV.
    } \label{fig:sfig10}
\end{figure}

\begin{figure}[H]
    \centering
    \includegraphics[width=1\columnwidth]{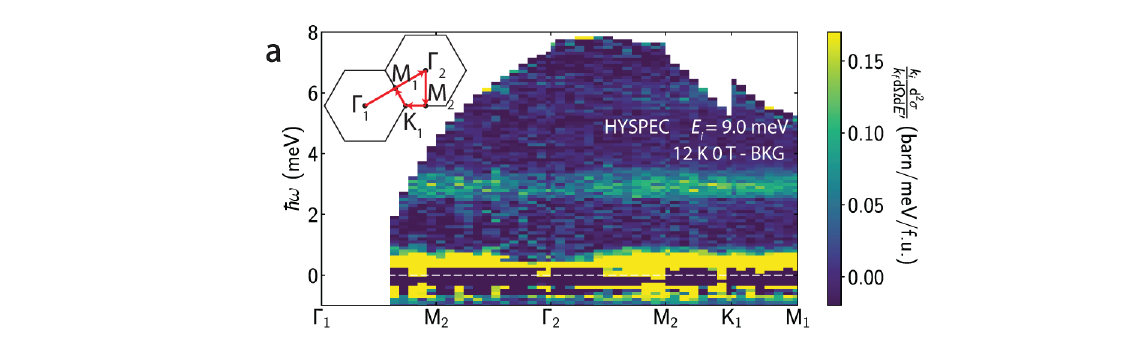}
    \caption{Magnetic neutron scattering cross section in the paramagnetic phase of  $\rm K_2Co(SeO3)_2$ at $T=12$~K versus $\hbar \omega$ and for $\textbf{Q}$ along high-symmetry directions. 
    The data were averaged along $l$, and the averaging window in the ab-plane is $\pm0.15/{\rm \AA}$ perpendicular to the high-symmetry direction.
    Low temperature 7 T data were subtracted as a background for $\hbar\omega<2.5$~meV while a constant was subtracted for higher energies. The incident neutron energy was $E_i = 9.0$~meV.
    } \label{fig:sfig11}
\end{figure}

\begin{figure}[H]
    \centering
    \includegraphics[width=1\columnwidth]{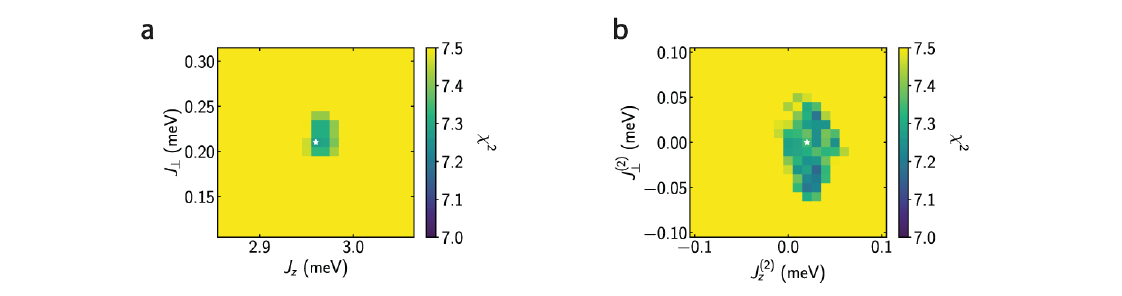}
    \caption{The $\chi^2$ goodness-of-fit parameter versus $J_{\perp,z}$ and $J_{\perp,z}^{(2)}$ for a pixel-to-pixel fit of the neutron scattering cross section associated with spin-wave excitations in the UUD plateau phase of $\rm K_2Co(SeO3)_2$ (Fig. 4a). From these fits, we infer $J_z=2.96(2)$~meV and $J_\perp=0.21(3)$~meV. Including anisotropic second nearest neighbor interactions $J_{\perp,z}^{(2)}$, the pixel-to-pixel analysis yields $J_{z}^{(2)} = 0.02(4)$~meV and $J_{\perp}^{(2)}=0.00(5)$~meV. The standard deviations for $J_{\perp,z}$ and $J_{\perp,z}^{(2)}$ were estimated by allowing 2\% increase of $\chi^2$.
    } \label{fig:sfig12}
\end{figure}

\begin{figure}[H]
    \centering
    \includegraphics[width=0.9\columnwidth]{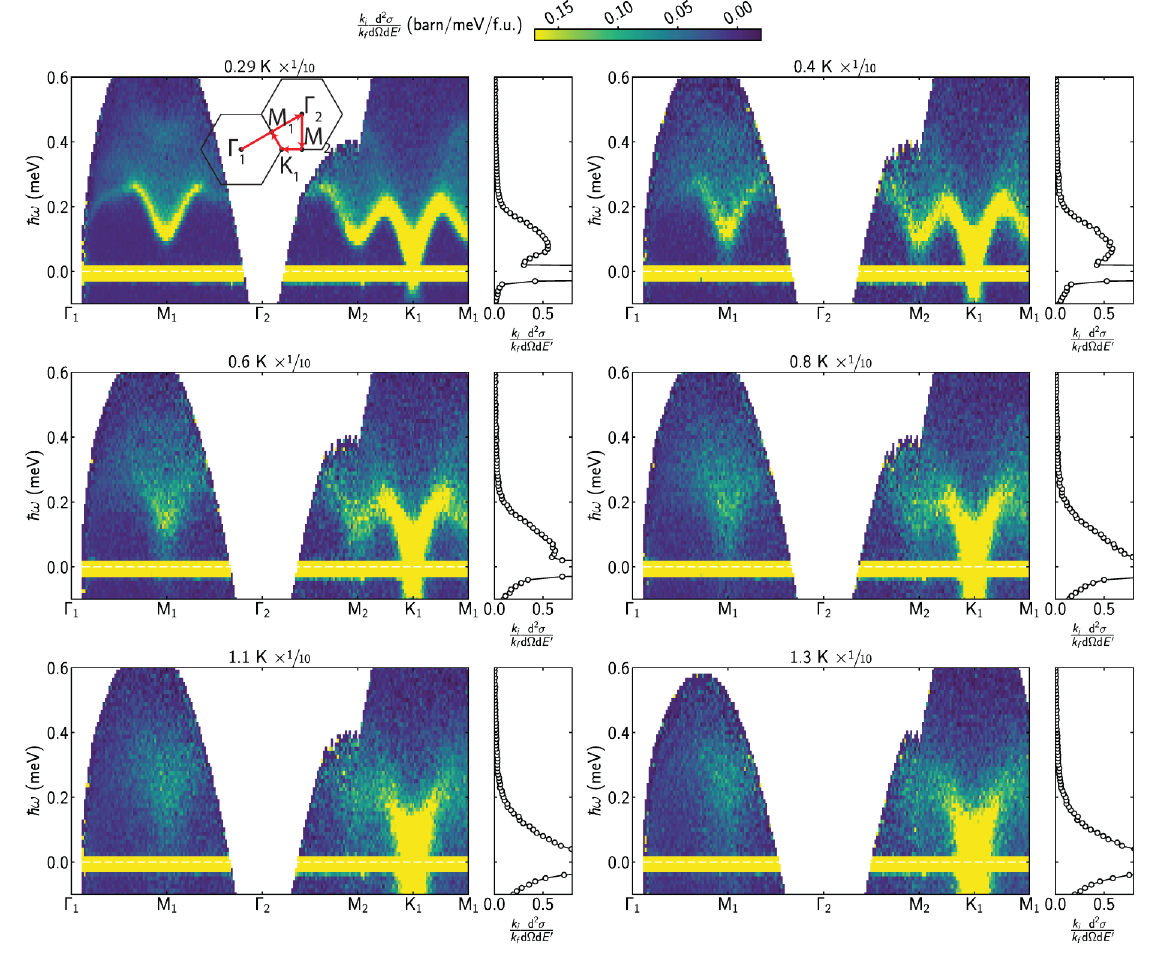}
    \caption{The inelastic neutron scattering cross section of $\rm K_2Co(SeO3)_2$ for wave vector transfer $\bf Q$ along high-symmetry directions. The right panels show constant-$\bf Q$ cuts versus $\hbar\omega$ at the K$_1$ point at various temperatures. The data were averaged along $l$, and the averaging window in the ab-plane is $\pm0.05/{\rm \AA}$ perpendicular to the high-symmetry direction. The data were acquired on CNCS with $E_i=1.0 $~meV in zero field.
    } \label{fig:sfig13}
\end{figure}

\begin{figure}[H]
    \centering
    \includegraphics[width=1\columnwidth]{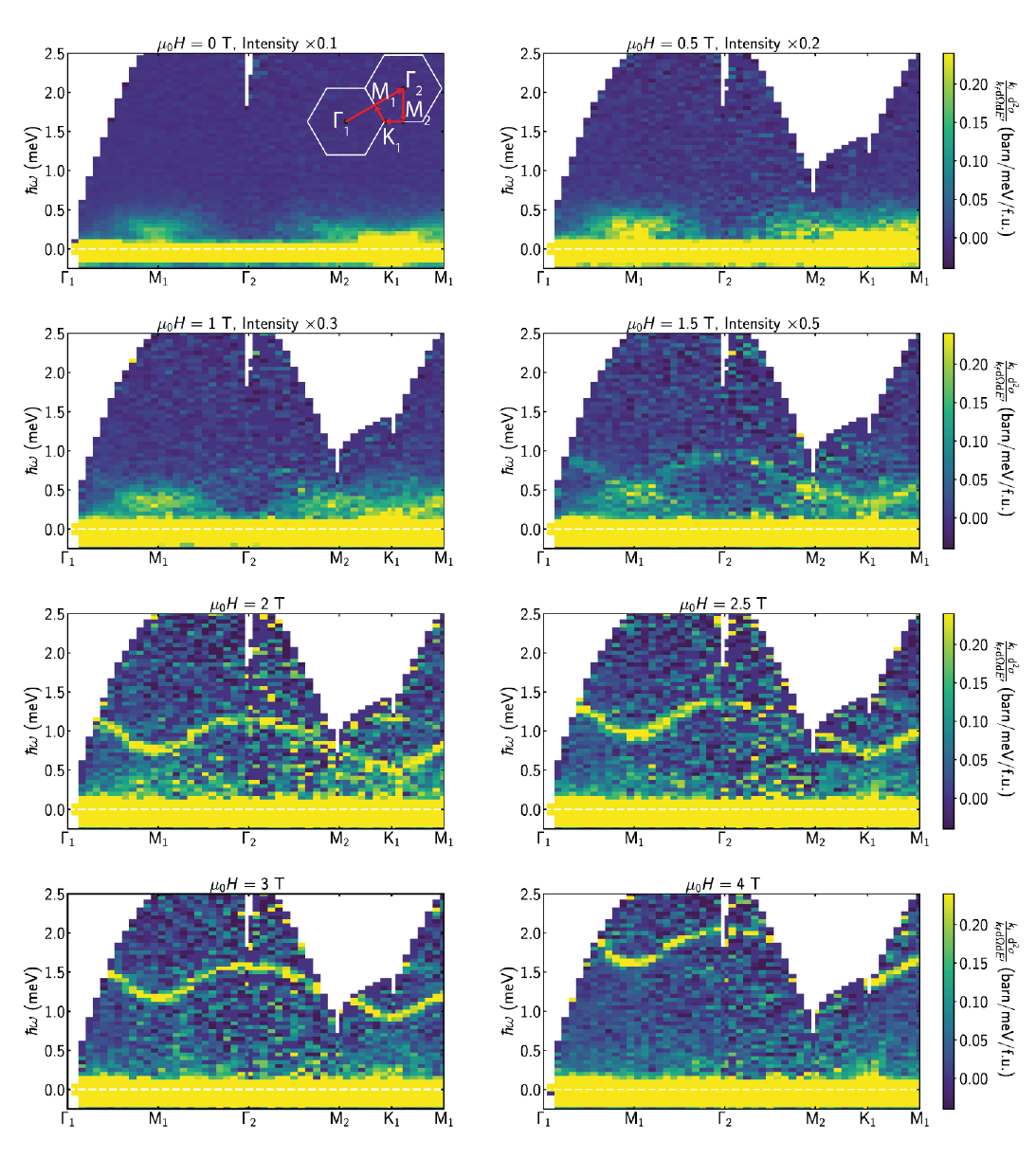}
    \caption{Field dependence of the inelastic magnetic neutron scattering cross section of $\rm K_2Co(SeO3)_2$ for wave vector transfer $\bf Q$ along high-symmetry directions. The data were averaged along $l$, and the averaging window in the ab-plane is $\pm0.15/{\rm \AA}$ perpendicular to the high-symmetry direction. The field was applied along the ${\bf c}$-axis, perpendicular to the horizontal scattering plane. The data were acquired at 1.8 K on HYSPEC with $E_i=3.8$~meV.  
    } \label{fig:sfig14}
\end{figure}

\section{\label{appendix: THz spectroscopy} THz spectroscopy}

A linearly polarized incident THz pulse is transmitted as elliptically polarized light. By measuring the final polarization state, we can determine both the diagonal ($T_{xx}$) and off-diagonal ($T_{xy}$) components of the complex transmission matrix ($T_{xx}$, $T_{xy}$). The complex Faraday rotation is defined as follows:

\begin{equation}
\theta + i\eta = \frac{T_{xy}}{T_{xx}}
\end{equation}

\begin{figure}[H]
    \centering
    \includegraphics[width=0.45\columnwidth]{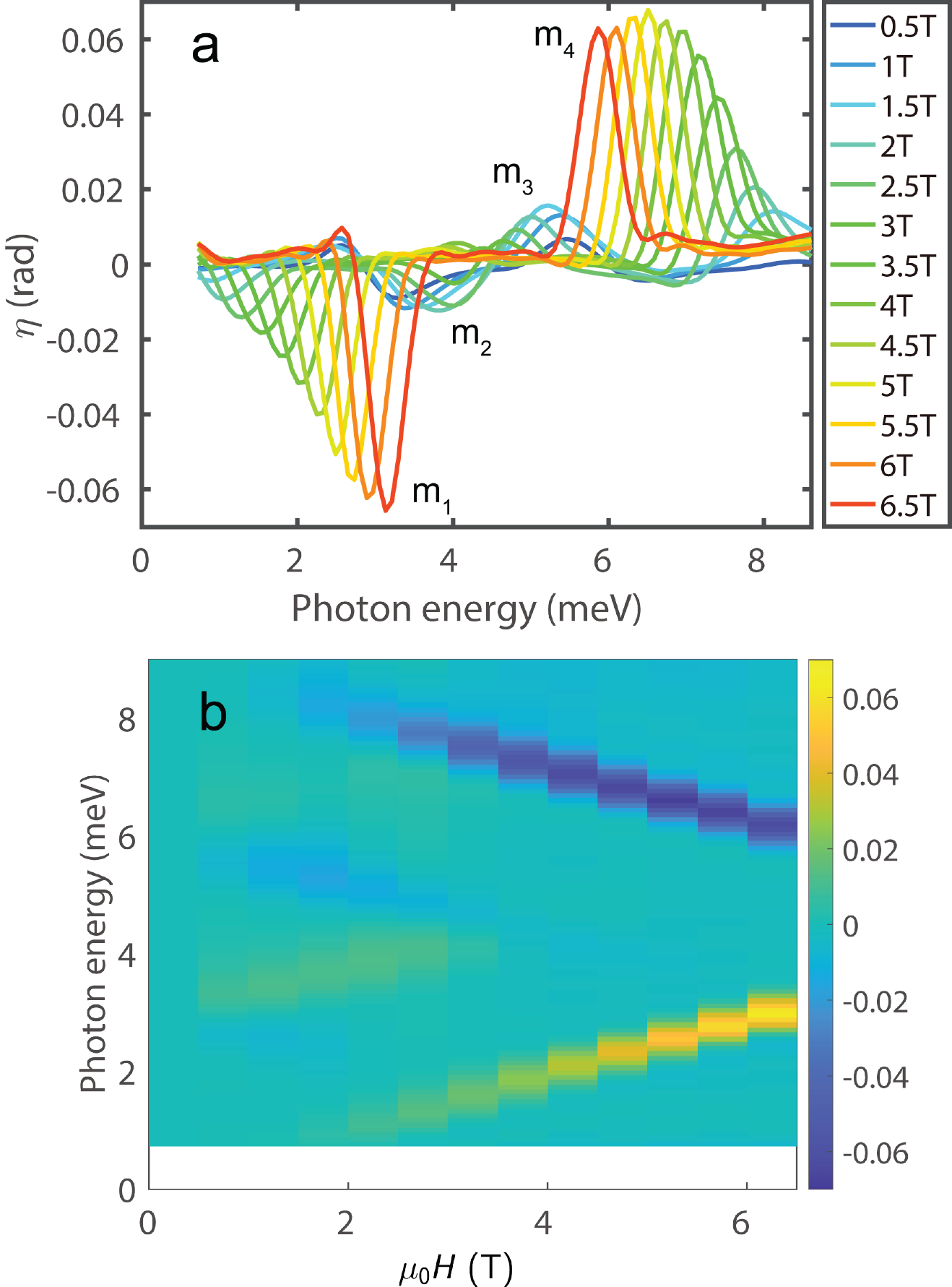}
    \caption{(a) Imaginary part of the Faraday rotation at various magnetic fields, with the magnetic field applied along the c-axis at 5~K.  (b) Color plot of the Faraday rotation.
    } \label{sfig:THz}
\end{figure}

This equation provides insight into the rotation angle ($\theta$) and ellipticity ($\eta$) induced by the Faraday effect. Figure~\ref{sfig:THz}a shows the Faraday rotation spectrum measured at 5~K. The Faraday rotation provides a signal relatively unaffected by non-magnetic absorption. The spectrum reveals distinct peaks in its imaginary part, corresponding to magnetic excitations.
Among these, we identify two prominent modes, labeled as m1 and m4, alongside two weaker modes, labeled as m2 and m3. The magnetic field dependence of these modes is illustrated in the 2D plots of Figure \ref{sfig:THz}b. Notably, all modes exhibit a linear dependence on the magnetic field. The sign of the Faraday rotation indicates the direction of magnetization. We observed that the signs of modes m1 and m4 are opposite, as are the signs of modes m2 and m3. This suggests that the two pairs of modes correspond to opposite spin processions. The direction of the magnetization is tuned by the applied magnetic field. If the applied magnetic field switches to the opposite direction, the sign of the rotations also switches.

\begin{figure}[H]
    \centering
    \includegraphics[width=0.45\columnwidth]{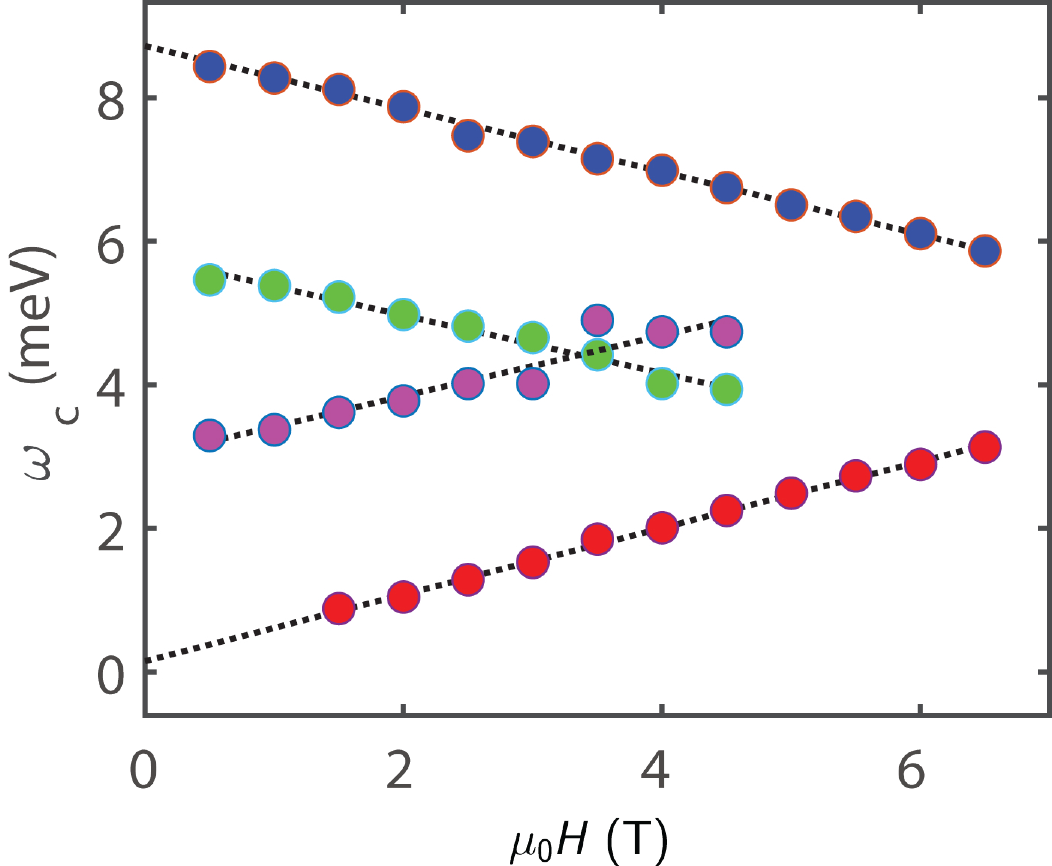}
    \caption{Frequencies for $Q=0$  ($\Gamma$) magnetic excitations in $\rm K_2Co(SeO_3)_2$ as functions of the applied c-axis-oriented magnetic field.
    } \label{sfig:THzBdep}
\end{figure}

We can determine the frequency and intensity of the modes from the imaginary part of the Faraday rotation. The evolution of the four modes with the magnetic field is shown in Figure~\ref{sfig:THzBdep}. All four modes shift linearly with the magnetic field. We fit the frequency evolution to the linear function :
\begin{equation}
    E(B) = E(0) - g_z \mu _{\rm B} \mu_0 H_z 
\end{equation}

The fitted values for the zero-field energy $E(0)$ and the $g$-factor $g_z$ for each mode are shown in the table:

\begin{center}
\setlength{\tabcolsep}{12pt}
\renewcommand{\arraystretch}{1.5}
\begin{tabular}{|c| c c c c| }

\hline
mode & m1 & m2 & m3 & m4 \\
\hline
$E$(0) (meV) & 0.16 & 2.99 & 5.78 & 8.73 \\
$g_z$ & -8 & -7.31 & 7.01 & 7.58 \\
\hline
\end{tabular}
\end{center}

In low fields, the four modes are related to the four bands of excitations observed with neutron scattering at $E_n(0) = n J_z$, $n = 0,1,2,3$. Upon increasing the magnetic field above 3~T, the system enters the UUD phase. In this phase, modes m2 and m3 vanish, while m1 and m4 evolve into spin-wave excitations and gradually shift to around 3~meV and 6~meV in the 7~T field. These results are again consistent with the neutron scattering spectra.

%